\documentclass[journal]{IEEEtran}
\ifCLASSINFOpdf
  \usepackage[pdftex]{graphicx}
  \graphicspath{../eps/}
  \DeclareGraphicsExtensions{.pdf,.jpeg,.png}
\else
  \usepackage[dvips]{graphicx}
  \graphicspath{{../eps/}}
  \DeclareGraphicsExtensions{.eps}
\fi
\usepackage{siunitx}
\usepackage[colorlinks=false]{hyperref}
\usepackage{color}
\usepackage{cite}
\usepackage{amsmath}
\usepackage{amssymb}
\usepackage{mathptmx}
\usepackage{epsfig}
\usepackage{times}
\usepackage{threeparttable}
\usepackage{stfloats}
\usepackage{amsmath,amssymb,amsfonts}
\usepackage{algorithmic}
\usepackage{textcomp}
\usepackage{array}
\usepackage{bm}
\usepackage{xcolor}
\usepackage{diagbox}

\usepackage{array}

  \usepackage[caption=false,font=footnotesize]{subfig}

\usepackage[colorinlistoftodos]{todonotes}
  \usepackage{fancyhdr}

\begin{document}
%
\title{PICO-RAM: A PVT-Insensitive Analog Compute-In-Memory SRAM Macro \\with In-Situ Multi-Bit Charge Computing \\and 6T Thin-Cell-Compatible Layout}
%
%

\author{Zhiyu Chen,~\IEEEmembership{Member,~IEEE,}
        Ziyuan Wen,~\IEEEmembership{Graduate Student Member,~IEEE,}
        Weier Wan, \\
        Akhil Reddy Pakala,~\IEEEmembership{Graduate Student Member,~IEEE,}
        Yiwei Zou,~\IEEEmembership{Graduate Student Member,~IEEE,}\\
        Wei-Chen Wei,
        Zengyi Li,
        Yubei Chen,
        Kaiyuan~Yang,~\IEEEmembership{Member,~IEEE}
\thanks{Manuscript received on} 
\thanks{Z. Chen, Z. Wen, A.R. Pakala, Y. Zou, W. Wei, and K. Yang are with the Department of Electrical and Computer Engineering, Rice University, Houston TX, 77005, USA.}
\thanks{W. Wan, Z. Li, and Y. Chen are with Aizip Inc., Cupertino CA, 95014, USA. Y. Chen is also with the Department of Electrical and Computer Engineering, University of California, Davis, Davis CA, 95616, USA.}
\thanks{(Corresponding Author: Kaiyuan Yang, kyang@rice.edu)}}

%
%

\markboth{Journal of Sold-State Circuits}%
{Shell \MakeLowercase{\textit{et al.}}: Bare Demo of IEEEtran.cls for IEEE Journals}
%


\IEEEpubid{\begin{minipage}{\textwidth}\ \\[12pt]
  \\
  \\
  \\
  \centerline{\copyright 2024 IEEE. Personal use of this material is permitted. Permission from IEEE must be obtained for all other uses,}\\ \centerline{in any current or future media, including reprinting/republishing this material for advertising or promotional purposes,} \\\centerline{creating new collective works, for resale or redistribution to servers or lists, or reuse of any copyrighted component of this work in other works.}
\end{minipage}} 



\maketitle

\begin{abstract}
Analog compute-in-memory (CIM) in static random-access memory (SRAM) is promising for accelerating deep learning inference by circumventing the memory wall and exploiting ultra-efficient analog low-precision arithmetic. Latest analog CIM designs attempt bit-parallel schemes for multi-bit analog Matrix-Vector Multiplication (MVM), aiming at higher energy efficiency, throughput, and training simplicity and robustness over conventional bit-serial methods that digitally shift-and-add multiple partial analog computing results. However, bit-parallel operations require more complex analog computations and become more sensitive to well-known analog CIM challenges, including large cell areas, inefficient and inaccurate multi-bit analog operations, and vulnerability to PVT variations. 
This paper presents PICO-RAM, a PVT-insensitive and compact CIM SRAM macro with charge-domain bit-parallel computation. It adopts a multi-bit thin-cell Multiply-Accumulate (MAC) unit that shares the same transistor layout as the most compact 6T SRAM cell. All analog computing modules, including digital-to-analog converters (DACs), MAC units, analog shift-and-add, and analog-to-digital converters (ADCs) reuse one set of local capacitors inside the array, performing in-situ computation to save area and enhance accuracy. A compact 8.5-bit dual-threshold time-domain ADC power gates the main path most of the time, leading to a significant energy reduction. Our 65-nm prototype achieves the highest weight storage density of 559 Kb/mm${^2}$ and exceptional robustness to temperature and voltage variations (-40 to 105 \textdegree C  and 0.65 to 1.2~V) among SRAM-based analog CIM designs.

\end{abstract}

\begin{IEEEkeywords}
CMOS; static random access memory (SRAM); compute-in-memory; mixed-signal computing; deep learning
\end{IEEEkeywords}

%
\IEEEpeerreviewmaketitle

\section{Introduction}
%
%
%
%

  \begin{figure}[t]
      \centering
      \includegraphics[scale=0.42]{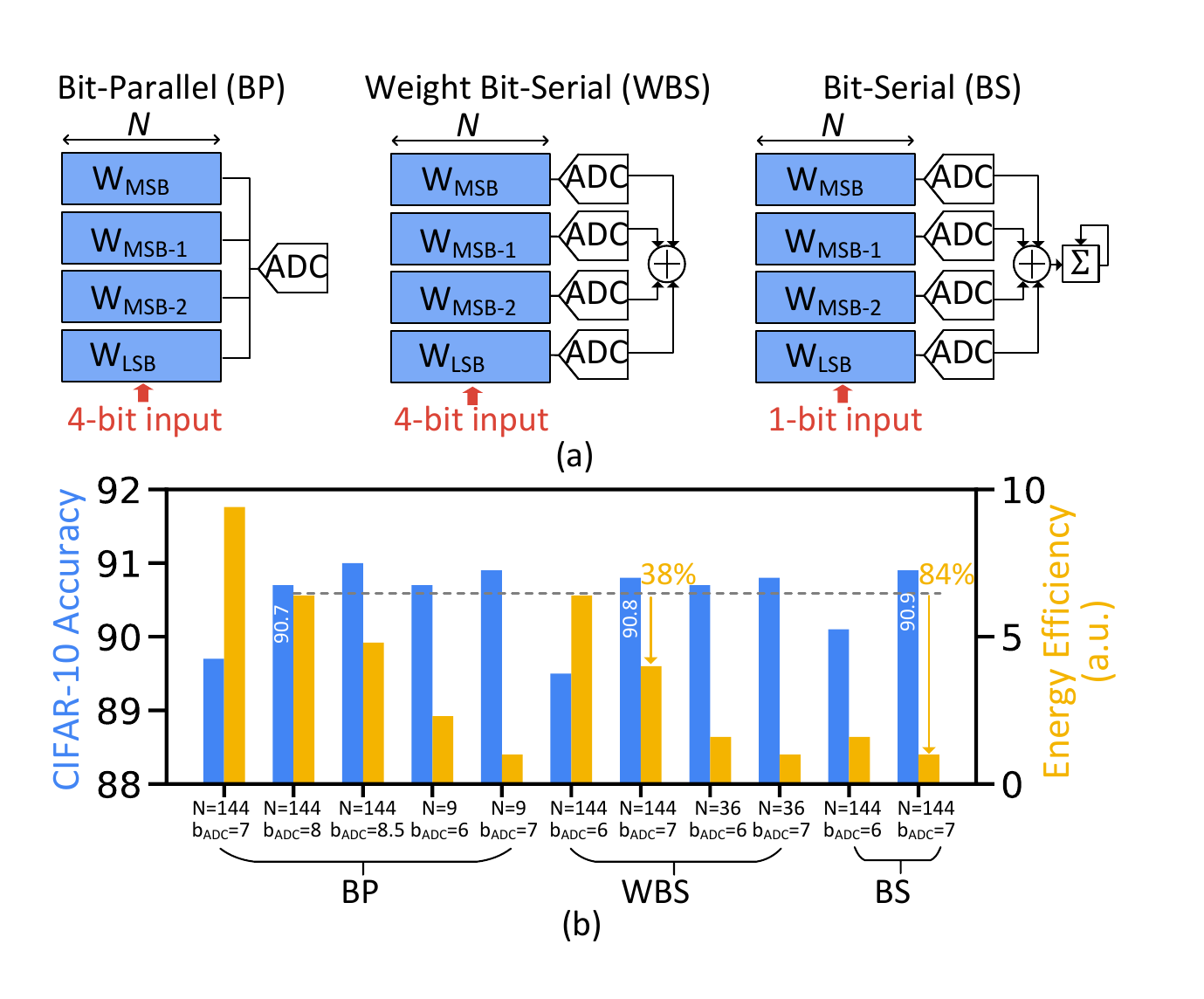}
      \vspace{-2ex}
      \caption{(a) BP, WBS, and BS schemes and (b) their simulated energy efficiency and CIFAR-10 accuracy (with ResNet-20) across CIM macro configurations.}
      \vspace{-2ex}
      \label{motivation}
   \end{figure}

\IEEEPARstart{N}on-Von Neumann compute-in-memory (CIM) has been a promising solution to circumvent the memory wall issues in data-intensive computation tasks~\cite{horowitzComputingEnergyProblem2014} such as deep neural networks (DNNs). CIM conducts MVM directly inside the memory with multiple rows of cells accessed concurrently, enhancing the data locality and amortizing the read energy~\cite{vermaInMemoryComputingAdvances2019}. Among various current and emerging memory technologies, SRAM, as a mature embedded memory, is of great interest due to its superior technology scalability and readiness, faster write/read speed, accurate weight storage, and simpler peripherals. While recent digital CIM in SRAM~\cite{DIMC_mingoo, tsmc_isscc22_1, tsmc_vlsi22, ECIM, wu202322nm, jedhea12,guo2023a28nm, chih1689TOPS162021, mori2023a4nm, kimColonnadeReconfigurableSRAMBased2021, he2023a28nm, yan20221} shows great potential in voltage scalability without sacrificing the accuracy, the adder tree and local logic gates significantly reduce the memory density and consume higher energy in low-precision models (4-8 bit) tailored for edge devices~\cite{murmannMixedSignalComputingDeep2021}. On the other hand, analog CIM takes advantage of ultra-efficient low-precision analog computation within a small cell footprint to achieve superior energy and area efficiency. A convolution in an analog CIM system can be expressed by: 
\begin{equation}
    Y=\mathbb{Q}(\sum_i W_iX_i)
    \label{bit_parallel}
\end{equation}

where $W_i$ and $X_i$ represent the weights and activations, respectively. $\mathbb{Q}(\cdot)$ denotes the non-ideal mapping from the mixed-signal computing. The conventional current-domain CIM turns on multiple wordlines simultaneously and accumulates current on bitlines~\cite{dong15351TOPS3722020, gonugondla42pJDecision12TOPS2018, yinxnor-sram, okumuraTernaryBasedBit2019, si1528nm64Kb2020, siTwin8TSRAMComputationinMemory2020, zhangInMemoryComputationMachineLearning2017}. It features a simple cell structure but suffers from PVT variations and nonlinear I-V characteristics of transistors. Recent charge-domain CIM SRAM macros~\cite{biswas2019conv, jiangC3SRAMInMemoryComputingSRAM2020, jiaProgrammableHeterogeneousMicroprocessor2020, jiaScalableProgrammableNeural2021, dctram, capram, leeChargeDomainScalableWeightInMemory2021,wangChargeDomainSRAM2023, hsieh2023a70, su1628nm384kb2021, hsuHighThroughputEnergyAreaEfficient2021, valavi64Tile4MbInMemoryComputing2019, chen2023a22nm, 9896828} have greatly improved the linearity of analog MVM and their robustness to process variations, approaching the inference accuracy of digital hardware in practical computer vision benchmarks.

However, there remain critical limitations in existing analog CIM SRAM designs regarding accuracy and efficiency for multi-bit MVM. Due to the binary nature of SRAM cells, they cannot natively support mult-bit MVM and rely on additional processing schemes to overcome the limitation. To simplify the analog MAC circuitry and increase the computing accuracy, many prior studies employ a \textbf{Bit-Serial (BS)} scheme, as shown in Fig.~\ref{motivation}. BS scheme can be formulated as:
 \begin{equation}
     Y=\sum_p^{B_W}\sum_q^ {B_X} 2^p 2^q  \mathbb{Q}(\sum_i W_i^pX_i^q)
 \label{bit_serial}
\end{equation}

where $W_i^p$ and $X_i^q$ represent the one-bit weights and activations involved in analog MACs, respectively. $B_W$ and $B_X$ are the actual bit precision of weights and activations in a quantized DNN model. BS indeed decreases the ratio between the possible analog MAC output levels and ADC quantization levels for each MVM. However, this ratio is not a reliable metric to evaluate accuracy because the subsequent digital accumulation of errors is ignored. In fact, since $ \mathbb{Q}(\cdot)$ is placed within the shift-and-add loops rather than outside, Equation~\ref{bit_serial} only changes the distribution of the MAC computing errors due to quantization noise, rather than reducing the range of errors as we can achieve with increasing ADC resolution. As detailed in Section~\ref{sec:sqnr}, we observe that such error distribution reshaping effect is less effective for enhancing the accuracy of analog MVM and deep learning inference, compared to directly enhancing ADC resolution or reducing the maximum number of analog MACs before ADC (see $N$ in Fig.~\ref{motivation}(a)). Meanwhile, the BS scheme increases energy consumption linearly with the analog MAC's input and weight bit precision, making it the least energy-efficient choice for enhancing the computing resolution. Finally, BS processing complicates the network training for analog CIM leading to unstable training dynamics and longer training time, because of the necessity to approximate the gradients for bit-wise MAC and to model quantization errors at the bit level~\cite{jin2022pim}.

\textbf{Bit-Parallel (BP)} CIM, on the other hand, computes the exact multi-bit analog MVM in Equation~\ref{bit_parallel}, without additional processing after quantization. BP achieves better energy efficiency at the same SQNR compared to BS. It also requires fewer modifications to software training and is more compatible with existing quantization methodology. The primary challenge with BP CIM lies in its circuit implementation, which we explain in Section~\ref{sec:sqnr} and address in this work. 
Due to the challenges of realizing BP CIM, many prior arts employ an intermediate \textbf{Weight Bit-Serial (WBS)}~\cite{capram, leeChargeDomainScalableWeightInMemory2021} scheme. WBS parallelizes the input bits using DACs but still requires serial processing for the weights. It partially alleviates the aforementioned issues of BS, achieving superior energy and area efficiency~\cite{leeFullyRowColumnParallel2021}, but is still sub-optimal at iso-accuracy. The simulations in Fig.~\ref{motivation}(b) illustrate the energy efficiency of BP is roughly 1.6$\times$ and 6.4$\times$ higher than that of WBS and BS, respectively, at similar inference accuracy on CIFAR-10 dataset using 4-bit ResNet-20.

Recent attempts of BP CIM macros~\cite{hsieh2023a70, wangChargeDomainSRAM2023, dong15351TOPS3722020} have demonstrated promising performance, but the implementations need to be optimized for efficiency and computing accuracy. To achieve analog shift-and-add, prior studies add extra weighted capacitors in the periphery~\cite{dong15351TOPS3722020} or near memory cells~\cite{hsieh2023a70, wangChargeDomainSRAM2023}. However, it is difficult to achieve satisfactory capacitor matching due to limited layout space, leading to degraded computing accuracy. Meanwhile, the local multi-bit analog MAC circuitry incurs significant area overhead. Another challenge is implementing an efficient and accurate DAC for multi-bit inputs. Previous capacitor-based DACs require power-hungry analog buffers~\cite{leeChargeDomainScalableWeightInMemory2021, hsieh2023a70} while the current-steering DACs are sensitive to PVT variations~\cite{capram}. 
Overall, an ideal BP CIM design should encompass a compact cell array and periphery, achieving multi-bit MVM with high accuracy and PVT robustness, and the elimination of power-hungry analog buffers in CIM macros. To the best of our knowledge, there has not been a solution that meets all these requirements.  

In this paper, we present PICO-RAM, a \textbf{P}VT-\textbf{I}nsensitive and \textbf{CO}mpact CIM SRAM macro that satisfies all demands, by exploiting four design ideas: (1) a charge-domain 4-bit MAC unit with 6T-thin-cell-compatible layout;  (2) an accurate in-situ charge-domain shift-and-add circuit, which improves throughput and efficiency, reduces area, and simplifies DNN training for CIM; (3) a PVT-insensitive in-situ capacitive DAC (C-DAC) without analog buffers, which improves accuracy and saves area; and (4) a compact and low-power dual-threshold time-domain ADC with power gating of the continuous comparator and D-flip-flops (DFFs). The rest of the article is organized as follows. Section~\ref{sec:sqnr} analyzes the advantages and challenges of BP CIM.  Section~\ref{sec:core} presents the key concepts and implementation of the core in-situ CIM circuits. Section~\ref{sec:adc} covers the ADC design. Section~\ref{sec:measurement} provides measurement results, followed by a conclusion in Section~\ref{sec:conclusion}.

\section{Prospects and Challenges of Analog BP CIM}
\label{sec:sqnr}

In this section, we provide a semi-empirical analysis of the SQNR of analog CIM macros and deep learning training process to prove the advantages of BP. We also outline the circuit design challenges for realizing analog BP CIM.

\subsection{SQNR Analysis}
SQNR has been widely adopted to evaluate the accuracy of quantized deep learning models and CIM systems~\cite{lin2016fixed, gonugondla2020fundamental}. Since our primary goal is to compare different computing schemes, we assume ideal circuit components and focus on the quantization errors. We empirically simulated the SQNR of different hardware configurations. The SQNR of a CIM-based deep learning system is defined by~\cite{jiaProgrammableHeterogeneousMicroprocessor2020}:
\begin{equation}
    \mathrm{SQNR}=\frac{\sum y^2}{\sum(y-\bar{y})^2}
\end{equation}

Here, $y=\sum_i^{K} W_iX_i$ where $W_i$ and $X_i$ are 4-bit integers randomly sampled from truncated Gaussian distribution. $K=R\times R\times C$ where R is the kernel size and C is the input channels. $\bar{y}$ is the quantized output from CIM following the exact computing flows for different schemes as described in Equation~\ref{bit_parallel} and~\ref{bit_serial}. We further consider partial sum accumulation for multiple macros when $K$ exceeds $N$. Each SQNR data point is obtained with a million Monte-Carlo simulations.

To study the relation between accuracy and energy, we assume a simple energy model for a complete analog MVM as follows,

\begin{equation}
    E_{MVM} = \frac{K}{N}\cdot\frac{B_A}{b_A}\cdot(\frac{B_W}{b_W}E_{ADC}+B_W NE_{MAC})
\end{equation}

where $b_A$ and $b_W$ represent the bit precision of inputs and weights involved in the analog MVM, respectively. $E_{ADC}$ is the energy of one ADC, which scales linearly with the resolution according to Walden's FoM~\cite{walden1999analog}. $E_{MAC}$ denotes the energy of each analog MAC for $b_A$-bit inputs and 1-bit weights. We assume $E_{MAC}$ does not scale with $b_A$ because driver-free DACs consume negligible energy~\cite{capram, dctram}. For the BP scheme, the subsequent $b_W$-bit analog shift-and-add also causes almost zero energy overhead (see details in Section~\ref{sec:isa}). Based on measurement results in~\cite{capram}, we set the energy ratio $E_{ADC}/(NE_{MAC})=3.0$ when the ADC is 7-bit and $N=144$.

   \begin{figure}[t]
      \centering
      \includegraphics[scale=0.35]{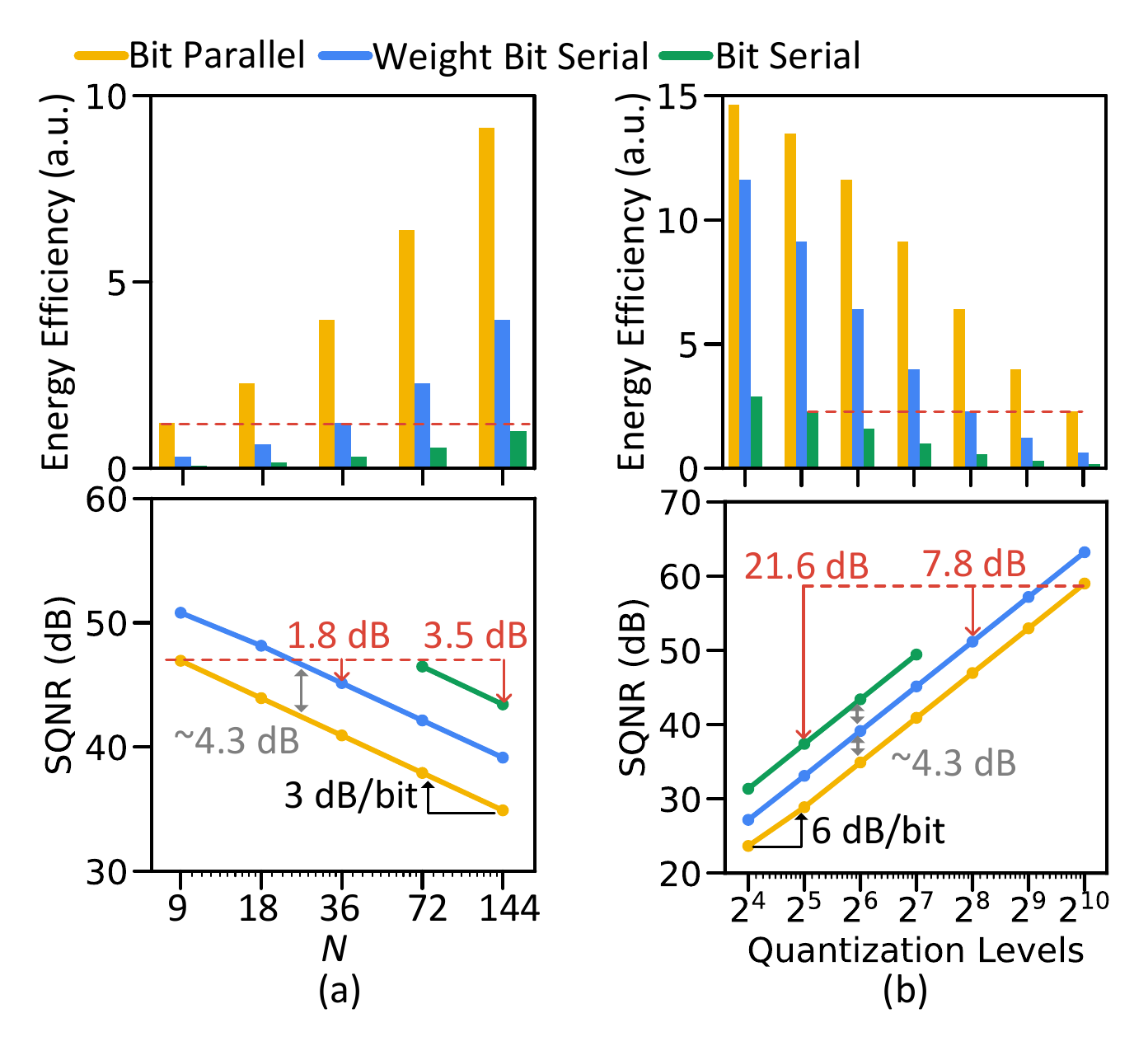}
      \vspace{-2ex}
      \caption{Simulated SQNR and energy efficiency under different hardware configurations when (a) quantization level = 64 and (b) $N$ = 144.}
      \label{sqnr}
   \end{figure} 

In Fig.~\ref{sqnr}(a), we fix the quantization level to 64 and sweep $N$  in each macro. At roughly the same energy efficiency, BP ($N=9$) achieves 1.8 dB and 3.5 dB SQNR improvement compared to WBS ($N=36$) and BS ($N=144$), respectively. Furthermore, Fig.~\ref{sqnr}(b) fixes N=144 and sweeps the quantization levels. BP increases SQNR by 7.8 dB and 21.6 dB at the same energy efficiency where quantization levels = 1024, 256, and 32 for BP, WBS, and BS, respectively. In summary, depending on the implementation details, BP achieves 3.5$\sim$21.6 dB higher SQNR than BS at the same energy efficiency. The SQNR results are highly correlated with the CNN inference accuracy in Fig.~\ref{motivation}, which validates the superior accuracy of BP.

It is important to note that different hardware configurations impact SQNR differently, even when they modify the analog-to-quantization level with the same factor. For example, while adding 1-bit ADC resolution increases SQNR by 6 dB, halving $N$ only increases 3 dB. Changing from WBS to 4-bit BP only decreases 4.3 dB, despite the analog dynamic range being amplified by 15$\times$. This is because for the same CNN model, adopting BS/WBS or reducing $N$ requires additional digital accumulation for the partial sums after the ADC. It accumulates quantization errors at each digital operation, which offsets the benefits of the increased sensing margin. In general, increasing the bit precision of analog MAC has minimal impacts on accuracy while its energy efficiency scales almost linearly with precision. This observation highlights a strategic consideration: accepting a modest reduction in accuracy due to increased analog MAC precision in exchange for significant gains in energy efficiency.

\subsection{Challenges of Deep Learning Training}
A common practice for CIM designs is providing customized training for deep learning models to account for non-idealities~\cite{jin2022pim, 9085999, zhang2023reshape}. A critical advantage of the software-friendly bit-parallel method is fewer modifications to the standard training flow. For traditional quantized deep learning models, a Straight-Through Estimator (STE) is adopted to back-propagate the non-differentiable $\mathrm{round}(\cdot)$ function~\cite{zhou2016dorefa}. For a real input $r_i\in[0,1]$, the derivative of the quantized output with respect to the input is given by:
\begin{equation}
    \frac{\partial}{\partial r_i}(\frac{1}{2^k-1} \mathrm{round}((2^k-1) r_i ))=1
\end{equation}

where $k$ is the quantization precision. However, integrating bit-serial computing into training workflows requires significant adaptation, such as employing a Generalized Straight-Through Estimator (GSTE)~\cite{jin2022pim}. Specifically, for a given input $x$, we assume:
\begin{equation}
    \mathrm{d}~\mathrm{round}(x)=\xi x
\end{equation}

where $\xi$ is an empirical scaling factor. This assumption, which is much stronger than the original STE, is necessary to calculate the bit-wise gradients for BS and WBS designs. This approximation leads to degraded accuracy, extended training latency, and frequent converging failures. Various strategies, such as empirically tuning $\xi$ to adjust training dynamics, must be employed to mitigate these challenges. In contrast, BP only adds one ADC quantization step after each convolution, which can be simply incorporated into conventional quantization frameworks with the original STE assumption.

\subsection{Challenges of Analog BP CIM}

    \begin{figure}[t]
      \centering
      \includegraphics[scale=0.95]{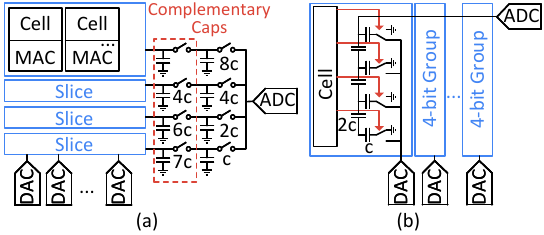}
       \vspace{-2ex}
      \caption{Prior charge-domain shift-and-add designs for analog BP CIM, using (a) peripheral weighted capacitors and (b) C-2C ladders.}
      \label{prior}
      \vspace{-2ex}
   \end{figure}

\textbf{\textit{Challenge 1}: Compact and accurate analog shift-and-add for multi-bit weights.} Prior studies add extra weighted capacitors in the periphery for charge-sharing across adjacent slices, as shown in Fig.~\ref{prior}(a)~\cite{dong15351TOPS3722020, hsieh2023a70}. However, it is difficult to achieve satisfactory matching among the weighted capacitors given the limited layout space, thereby leading to computing nonlinearity. Meanwhile, extra capacitors will share the charge from the analog MAC and therefore reduce the input voltage range of ADC. This disadvantage escalates when complementary capacitors are incorporated to maintain the same total capacitance among different slices. Another critical concern is that the ADC cannot utilize the local capacitors in the array as the sampling capacitor. The small peripheral capacitors increase thermal noise and potentially require additional analog buffers.

A recent work~\cite{wangChargeDomainSRAM2023} presented an in-array C-2C ladder for analog shift-and-add, as depicted in Fig~\ref{prior}(b). 
All the capacitors in the array can be reused as the sampling capacitor of ADC, ensuring the input swing of ADC is unaffected. However, the matching between the C and 2C capacitors becomes even more challenging given the extremely restricted local layout space and the influence of parasitic capacitors. Additionally, the inclusion of extra switches to local circuits significantly decreases overall memory density, leading to a 9T cell and a MAC unit that each occupies twice the area of a 6T cell. Lastly, input DACs must use a resistive divider to drive the C-2C ladder with constant power consumption.

\textbf{\textit{Challenge 2}: Driver-free and accurate DAC designs for multi-bit inputs.} In charge-domain CIM designs, the most straightforward way of delivering the DAC output voltage to local capacitors is utilizing an analog buffer or a resistive divider. However, given the large output load and speed requirement, the overhead for strong driving capability is significant but often overlooked by previous studies in their energy estimation where multiple off-chip reference voltages are supplied. As an example, \cite{leeChargeDomainScalableWeightInMemory2021} reports its DAC drivers occupy 11.4\% of the macro area and incur 94-pJ energy overhead in 28 nm, accounting for 68.5\% of the total energy in a macro that supports 5-bit activations and 8-bit weight. ~\cite{hsieh2023a70} adopts a simple two-transistor push-pull buffer to save energy, but the linear output range is highly restricted. On the other hand, some designs use a current-steering DAC to avoid power-hungry analog buffers~\cite{dctram, capram,biswas2019conv}. However, it is sensitive to PVT variations and thus requires a complicated calibration process in practice.

\textbf{\textit{Challenge 3}: Efficient and compact ADC designs over 8 bits.} Because of the multi-bit analog computation, BP typically has more analog levels before ADC and thus requires a higher resolution to obtain sufficient SQNR. The time-domain ADC offers a compact area, technology scalability, and sufficient conversion speed at this resolution range~\cite{dctram}. However, its energy linearly scales with the quantization levels, making it less energy-efficient than the conventional SAR ADCs.

\section{In-Situ Charge-Domain Computing with 6T Thin-Cell Layout}
\label{sec:core}


\subsection{Principles of the Core Circuits}

The key of the proposed ``in-situ'' capacitive computation is the recurrent usage over a single set of metal-oxide-metal (MOM) capacitors for all analog tasks, including DAC, analog MAC, analog shift-and-add and ADC, without extra peripheral circuitry. Throughout the entire analog processing chain, transistors only act as switches for fully charge-domain operations, eliminating PICO-RAM's sensitivity to PVT variations of transistors. This approach is crucial for reducing area, mitigating computing nonlinearity, and eliminating buffering and sampling circuits. Meanwhile, despite various capacitor configurations for different tasks, the overhead of the computing circuitry in the array is reduced to minimal since it adopts a 6T-thin-cell-compatible layout.

\begin{figure}[t]
      \centering
      \includegraphics[scale=0.7]{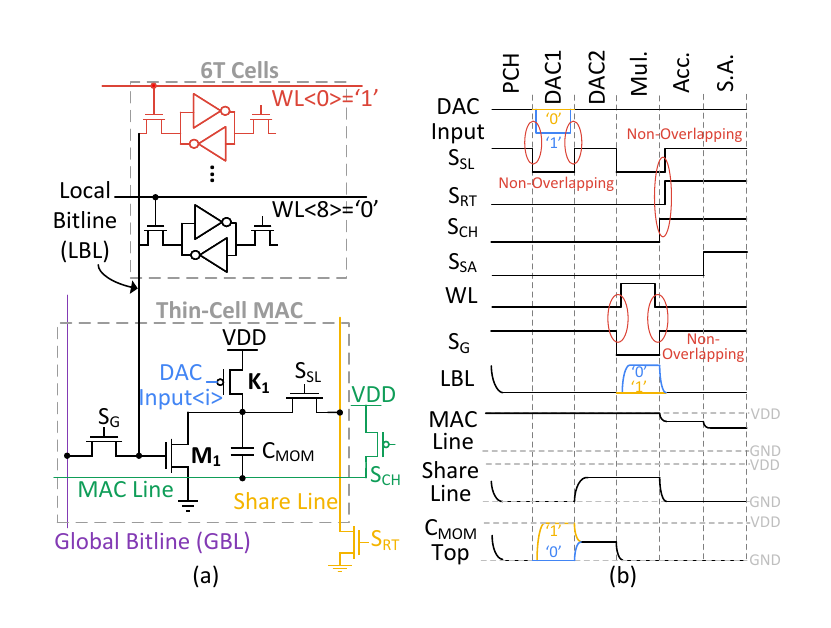}
       \vspace{-2ex}
      \caption{Proposed 6T-thin-cell-compatible cluster and operating waveforms.}
      \label{cluster}
   \end{figure}

\begin{figure}[t!]
      \centering
      \includegraphics[scale=0.75]{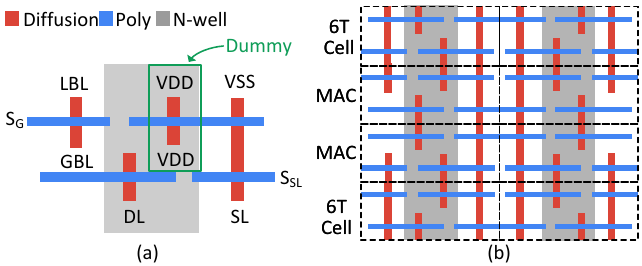}
      \vspace{-2ex}
      \caption{Thin-cell MAC unit layout and integration with 6T cells.}
      \label{layout}
   \end{figure}

As shown in Fig.~\ref{cluster}(a), the core building block of the PICO-RAM macro is an SRAM cluster. Each cluster consists of (1) nine standard 6T SRAM cells to store weights and (2) a thin-cell MAC unit performing multi-bit charge-domain MAC and configuring connections of the capacitor. The prechargers for Share Line and MAC Line, controlled by S${\rm _{CH}}$ and S${\rm _{RT}}$, are shared across columns vertically and rows horizontally. One of the nine cells will be accessed in each operation, while the rest of the inactive cells store weights from other layers or channels to improve storage density~\cite{capram}. Fig.~\ref{layout} illustrates the layout of the MAC unit. It shares the exact transistor layout as the most compact 6T SRAM cell, differing only in metal connections. With such a thin-cell cluster, the weight storage density may approach that of a commercial SRAM if the same push-rule layout is adopted, and the matching between transistors is also improved due to the regular layout. The MOM capacitor C${\rm _{MOM}}$s ($\sim$4fF) within the MAC unit is placed above the cluster to save area. The layout is further verified in 28 nm CMOS, achieving the same area as a 6T SRAM cell with an area of 0.27 $\mu$m$^2$. For simplicity, the wordline and bitline for the access transistors on the right side of the 6T, which are only used for normal read/write, are omitted in Fig.~\ref{cluster}(a).

\subsection{In-Situ Shift-and-Add}
\label{sec:isa}

   \begin{figure}[t]
      \centering
      \includegraphics[scale=0.7]{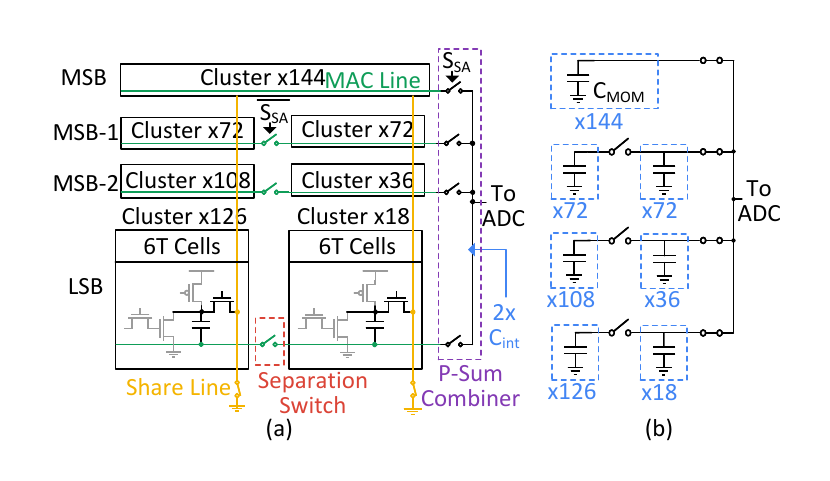}
      \vspace{-1ex}
      \caption{(a) Diagram of the in-situ shift-and-add circuits in a CIM MVM group and (b) connection of C${\rm _{MOM}}$s when shift-and-add occurs.}
      \label{shift_add}
   \end{figure}

    \begin{figure}[t!]
      \centering
      \includegraphics[scale=0.3]{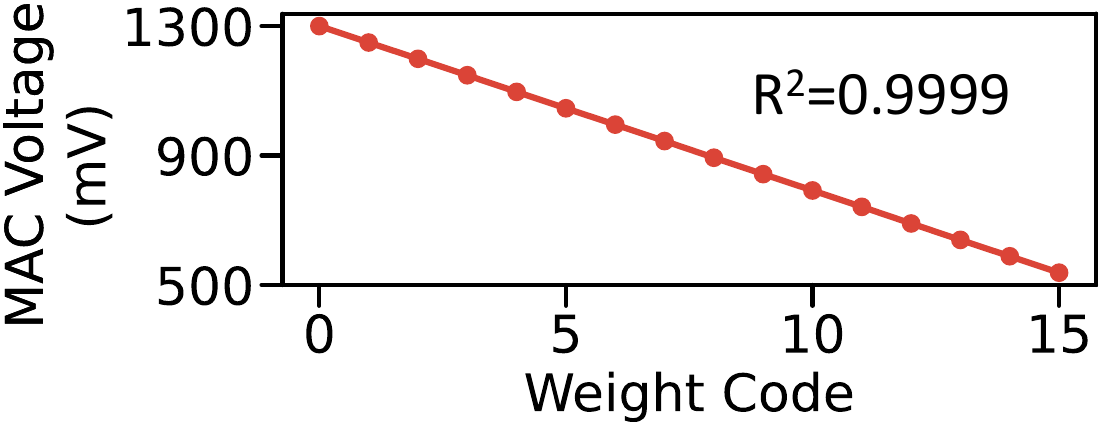}
      \caption{Post-layout linearity simulation of in-situ shift-and-add by sweeping 4-bit weight values in SRAM.}
      \label{linearity_sa}
   \end{figure}
To address \textit{Challenge 1}, the analog shift-and-add for multi-bit weights deploys in-situ C${\rm _{MOM}}$s in the cluster for weighted charge-sharing, resulting in superior compactness and computing accuracy, as illustrated by Fig.~\ref{shift_add}(a). In our implementation, 144 clusters integrate into a slice where their MAC Line is connected together. Inside the slices MSB-1, MSB-2, and LSB, separation switches are inserted to disconnect MAC Lines. The number of clusters (72, 36, and 18) on the right side of the separation switch represents the bit's weight. The entire clusters (144 in total) in the MSB slice participate in the weighted summation. The shift-and-add happens right after the conventional charge-domain computation on MAC Line when the accumulation results are ready on C${\rm _{MOM}}$s. During this phase (S.A. phase in Fig.~\ref{cluster}(b)), S${\rm _{SA}}$ is high to turn off the separation switches and connect the neighboring 4 MAC Lines using Partial-Sum (P-Sum) Combiner, facilitating a charge-sharing shift-and-add. The final connection of C${\rm _{MOM}}$s is shown in Fig.~\ref{shift_add}(b), forming an inter-slice weighted capacitive adder. 

Unlike prior works that face challenges of matching weighted capacitance value using diminutive capacitors in/near memory, the uniformly placed C${\rm _{MOM}}$s in clusters naturally offer superior matching and combines into a large total capacitance that greatly alleviates the effects of parasitic capacitors. The post-layout linearity simulation in Fig.~\ref{linearity_sa} sweeps the weight value in the memory while keeping the same input value, which demonstrates great linearity with $R^2=0.9999$.
The extra switches adopt a thin-cell transistor layout similar to MAC units, leading to only 3.4\% area overhead.

\subsection{In-Situ C-DAC}
\label{sec:dac}

   \begin{figure}[t]
      \centering
      \includegraphics[scale=0.7]{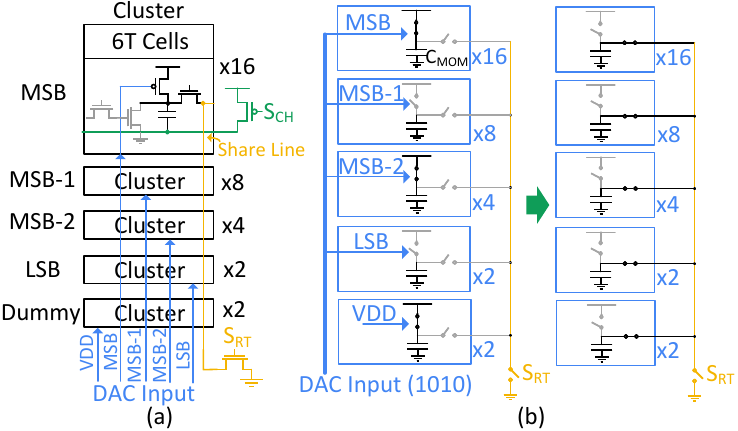}
      \vspace{-2ex}
      \caption{(a) Diagram of the in-situ C-DAC and (b) connection of C${\rm _{MOM}}$s in the two DAC phases.}
      \label{dac}
   \end{figure} 
   
 Leveraging a similar concept, the column-parallel 4-bit in-situ C-DAC (Fig.~\ref{dac}(a)) reuses C${\rm _{MOM}}$s as a capacitive voltage divider to address \textit{Challenge 2}. It is more robust to PVT variations than current-steering DACs and has a much smaller area overhead than the designs with explicit voltage dividers and analog buffers. Here, 32 clusters combine into a column with Share Line connected together. To realize the in-situ C-DAC, one memory column is divided into 4 groups, where switches K1 (see Fig.~\ref{cluster}(a)) in each group are controlled by a different bit from the 4-bit DAC input. The number of clusters in a group (16, 8, 4, and 2) represents the weight of the corresponding input bit. The conversion contains two phases of switching. During the first phase (DAC1 in Fig.~\ref{cluster}(b)), the top plates of C${\rm _{MOM}}$s are either pulled up to VDD if the bit is logic ‘1’ (0 V), or kept at zero if the bit is logic ‘0’ (VDD). In the DAC2 phase, the charge on C${\rm _{MOM}}$s is shared through the Share Line vertically with S${\rm _{SL}}$ set high and S${\rm _{RT}}$ set low. The output voltage is naturally sampled on C${\rm _{MOM}}$s for future computation. Therefore, no power-hungry analog buffer is needed after the DAC. An example of a capacitor connection in the two phases with a digital input of `1010' is in Fig.~\ref{dac}(b).

In addition to accuracy and area benefits, the two-phase in-situ C-DAC is also more energy efficient than conventional designs that employ analog buffers to directly drive the local C${\rm _{MOM}}$s. Meanwhile, C${\rm _{MOM}}$s perform as the voltage divider and the sampling capacitor simultaneously, further saving the charging energy for capacitors. The in-situ C-DAC is also aware of input sparsity since the capacitors will not be charged when inputs are zero. Our measurement demonstrates that the DAC only occupies 2.4\%$\sim$14.6\% of the total energy, depending on the input sparsity.

A few prior studies reported a similar strategy that reuses in-array capacitors as a reference generator for DACs~\cite{kim2022charge} or SAR-ADCs~\cite{hsuHighThroughputEnergyAreaEfficient2021}. However, all of the existing designs require a complex structure for cells and MAC units, diminishing the area benefits. The proposed design is the first one that embeds the C-DAC into the memory with minimal overhead.

\subsection{End-to-End CIM Operations}

   \begin{figure}[t]
      \centering
      \includegraphics[scale=0.35]{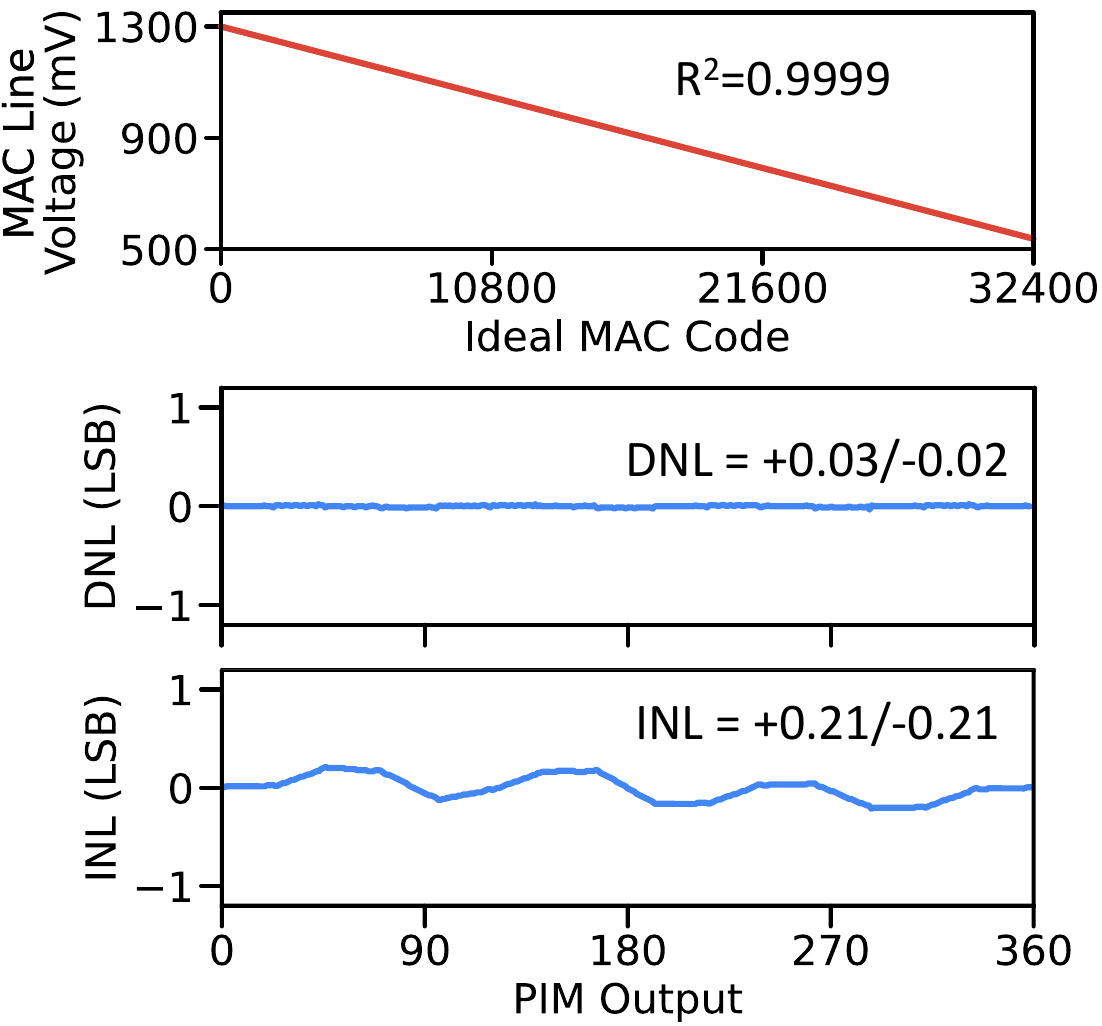}
      \vspace{-2ex}
      \caption{Simulated linearity of the charge-domain computing from the digital input to the MAC Line output.}
      \label{simulated_linearity}
   \end{figure} 
   
The CIM operation starts with a precharge phase (PCH phase), as shown in Fig.~\ref{cluster}(b). During this phase, the top plates of C${\rm _{MOM}}$s, MAC Lines, and Share Lines are initialized to ground, VDD, and ground, respectively. Then the in-situ C-DAC takes advantage of all the C${\rm _{MOM}}$s in a column, functioning as a reference generator, and samples the output voltage on the C${\rm _{MOM}}$s' top plates, while their bottom plates are grounded via MAC Lines (DAC1 and DAC2 phases). Subsequently, one of the WLs is activated to engage the NMOS M1 for the multiplication process (Mul. phase). Depending on the data stored in the 6T cell, the C${\rm _{MOM}}$s either discharge entirely or maintain their DAC voltages. During the accumulation phase (Acc. phase), S${\rm _{SL}}$ and S${\rm _{RT}}$ are set to a high state, grounding the top plates, and causing charge sharing across the C${\rm _{MOM}}$s connected to the same MAC Line in a given row. The in-situ charge-domain shift-and-add, enabled by S${\rm _{SA}}$, yet again employs local C${\rm _{MOM}}$s and conducts weighted charge-sharing across neighboring rows (S.A. phase). After the analog CIM, the TD-ADC reuses C${\rm _{MOM}}$s once more for voltage sampling and charge integration. The post-layout simulation in Fig.~\ref{simulated_linearity} demonstrates great linearity of the end-to-end charge-domain CIM with R$^2$=0.9999. Assuming an 8.5-bit quantization system, the DNL and INL of the transfer curve are bounded within +0.03/-0.02 and +0.21/-0.21 LSB, respectively. 

\subsection{Support of Signed Weights}
While PICO-RAM primarily conducts unsigned analog computation only, it can support signed arithmetic with a new encoding scheme. We map the signed 4-bit data from -8 to 7 into unsigned data from 0 to 15 and store them in the memory. After the unsigned analog computation, we subtract the offset to obtain the final signed results. Specifically, the original bit-parallel Equation~\ref{bit_parallel} can be re-formulated as:
\begin{equation}
    Y=\mathbb{Q}(\sum_i \tilde{W_i}X_i) - 8\sum_i X_i
    \label{signed_mac}
\end{equation}

where $\tilde{W_i}$ is the 4-bit unsigned weight after our mapping scheme. $\sum_i \tilde{W_i}X_i$ is calculated using the analog CIM macros while the summation of inputs $\sum_i X_i$ can be implemented with a simple digital adder tree. Note that in a real CIM system, a single adder tree can be shared among not only different slices but also numerous CIM macros. Therefore, its overhead is negligible. Compared to adopting specialized signed analog logic~\cite{wang2023a28nm,yinxnor-sram, siTwin8TSRAMComputationinMemory2020}, this method is more flexible, more accurate, and applicable to most CIM architecture.

 \section{Dual-Threshold Time-Domain ADC}
 \label{sec:adc}

   \begin{figure}[t]
      \centering
      \includegraphics[scale=0.36]{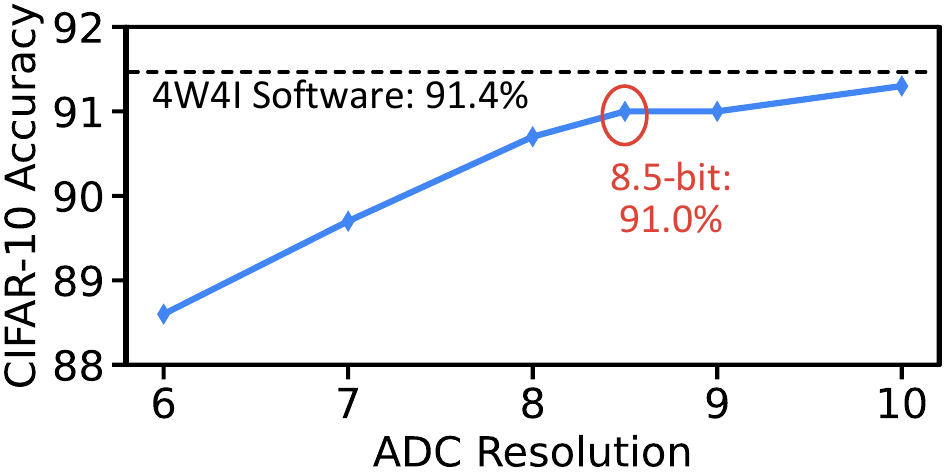}
      \vspace{-1ex}
      \caption{Simulated CIFAR-10 accuracy on 4-bit ResNet-20 with different ADC resolutions. }
      \vspace{-2ex}
      \label{simulated_inference}
   \end{figure} 
   
When designing CIM macros, one of the most critical considerations is selecting the ADC resolution. Due to the area and energy constraints, it is prohibitive to have a full-precision quantization. For example, PICO-RAM needs a 15-bit ADC to cover every level of the analog inputs. Therefore, recent CIM studies always allow a discrepancy between the quantization levels and the actual analog levels to maximize energy efficiency because the quantization errors may be tolerated by deep learning models. This is one of the key reasons why analog CIM achieves superior energy efficiency. For instance, ~\cite{hsieh2023a70, wang2023a28nm} select the ratio of the analog-to-quantization levels as high as 16178 while ~\cite{jiaScalableProgrammableNeural2021} chooses 4.5. Many studies lie between these extremes, such as ~\cite{siTwin8TSRAMComputationinMemory2020} (ratio=52) and~\cite{capram} (ratio=15). However, there is no general quantitative methodology for analyzing optimal ADC resolution because of different SQNR requirements, deep learning models, and datasets. 

We empirically find the suitable ADC resolution for PICO-RAM based on a CIM-Aware deep learning framework~\cite{jin2022pim}. It considers CIM characteristics such as additional ADC quantization, bit-serial computation, and analog non-idealities during training and generates corresponding inference accuracy. As shown in Fig.~\ref{simulated_inference}, we sweep the ADC resolution while fixing other hardware configurations of PICO-RAM. For the target CIFAR-10 dataset with 4-bit ResNet-20, the inference accuracy starts saturating at 8-bit ADC. In this case, the SQNR is high enough to approximate the software baseline. Therefore, we select an 8.5-bit ADC to achieve the best accuracy and energy trade-off. 

   \begin{figure}[t]
      \centering
      \includegraphics[scale=0.8]{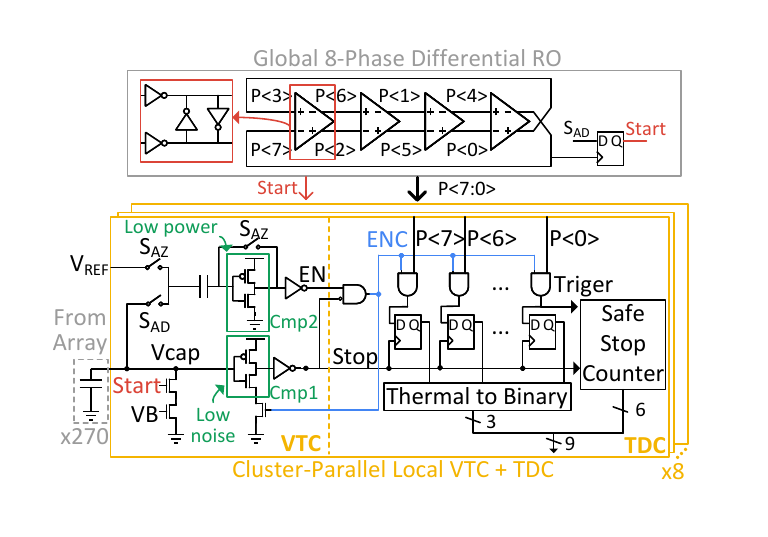}
      \vspace{-1ex}
      \caption{Diagram of the dual-threshold TD-ADC. }
      \label{adc}
   \end{figure} 
   \begin{figure}[t]
      \centering
      \includegraphics[scale=0.7]{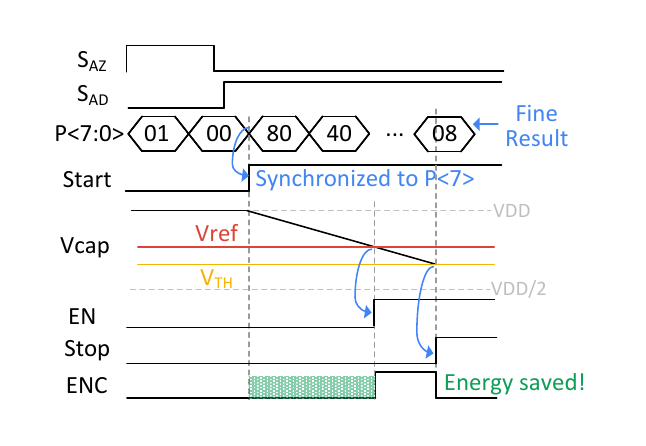}
       \vspace{-1ex}
      \caption{Operational waveforms of the dual-threshold TD-ADC. }
      \vspace{-2ex}
      \label{adc_wv}
   \end{figure} 

To address \textit{Challenge 3}, we design a dual-threshold TD-ADC architecture for better energy efficiency while maintaining the technology scalability and compact area. The 8.5-bit TD-ADC includes a voltage-to-time converter (VTC), a TDC, and a shared 8-phase differential ring oscillator (RO), as shown in Fig.~\ref{adc}. The VTC discharges the capacitors attached to MAC Lines until it reaches the threshold voltage of the zero detector (Cmp1), converting output voltage (Vcap) into a pulse. Thanks to the in-situ shift-and-add mechanism, the integration capacitor of the VTC is the combination of C${\rm _{MOM}}$s from four slices. The total capacitance is almost doubled over a bit-serial counterpart, significantly reducing the thermal noise and the current source noise from the VTC. To avoid exponentially increased area in conventional flash TDC, we adopt a compact folding-flash TDC topology~\cite{dctram}. The local registers sample the phases of the RO to generate the 3b fine results and the local counter triggered by one of the phases in the RO generates the 6b coarse results. The RO is free running to avoid a long settling time while synchronized to the ADC start signal S$\rm _{AD}$ to prevent an uncertain initial state (see Fig.~\ref{adc_wv}). The safe-stop mechanism synchronizes the counter's Stop and Trigger signals, preventing possible MSB errors caused by a wrong count when the two signals collide~\cite{dctram}.

Our TD-ADC features superior voltage scalability (down to 0.65 V) and ultra-compact area. With a shared RO, TD-ADC occupies 387.9 $\mu$m$^2$ each, overall (8 ADCs) accounting for only 4.6\% of the macro’s area. Sharing the RO also benefits the phase noise and linearity since the stage delays can be up-sized with few area and energy concerns. The local registers that dominate the TDC area utilize a custom true single-phase clocked (TSPC) structure which is 65\% smaller than a standard-cell DFF, leading to further area reduction.

One critical concern of this design is the high energy consumption throughout the conversion. The continuous zero detector (Cmp1) in the VTC must spend high power to suppress noise and the transparent dynamic latches in the TSPCs keep sampling the fast clock edges from the RO. To save power, a second low-power comparator (Cmp2) is added to power gate Cmp1 and TSPCs. Cmp2 is auto-zeroed by S$\rm _{AZ}$ before conversion, allowing it to maintain a low-power profile with near-minimum sizing while achieving minimal offset. It has a slightly higher threshold (set by Vref) than Cmp1 to disable the main path of ADC most of the time, leading to a 55.8
\% energy reduction of the local ADCs without compromising accuracy. 
\section{Measurement Results}
\label{sec:measurement}

   \begin{figure}[t]
      \centering
      \includegraphics[scale=0.7]{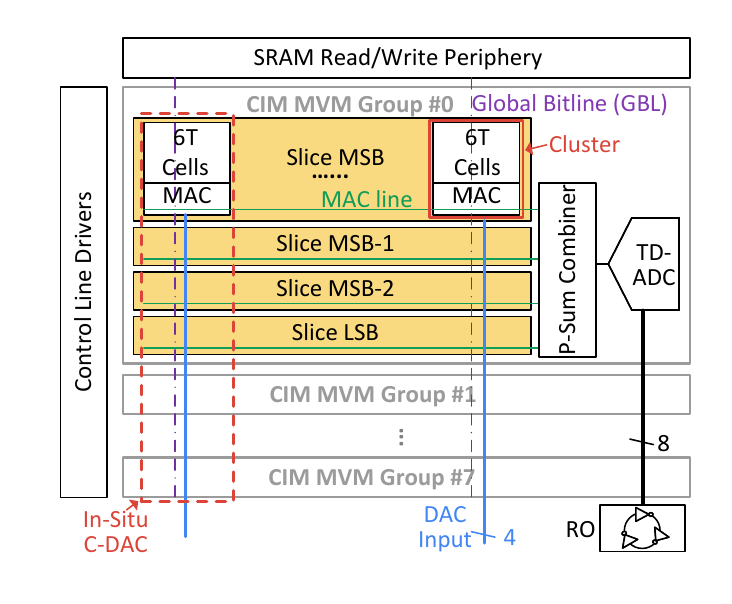}
      \caption{Macro diagram of the multi-bit in-situ CIM SRAM.}
      \label{macro}
    \end{figure}

    \begin{figure}[t]
      \centering
    \includegraphics[scale=0.9]{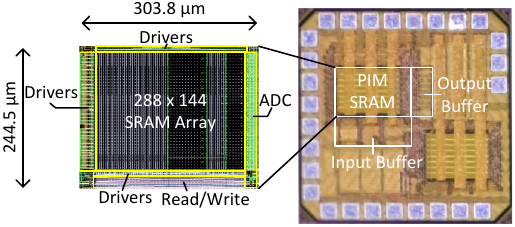}
      \caption{Die micrograph and macro layout.}
      \label{die}
   \end{figure}

Our prototype 288$\times$144 PICO-RAM macro contains eight CIM MVM groups, each connected to a TD-ADC as shown in Fig.~\ref{macro}. Within each group, four slices are present, conducting charge-domain vector multiplication with 4-bit activations and 4-bit weights, where each bit of the weights is stored in a corresponding slice. Each slice includes 144 clusters, each of which has nine 6T SRAM cells and a MAC unit. The macro completes 4-bit analog MVM in a single clock cycle, yet can support higher precision by leveraging the peripheral digital serial processing~\cite{jiaScalableProgrammableNeural2021,capram}.
 
The test chips are fabricated in 65-nm LP CMOS, as shown in Fig.~\ref{die}. The 40.5 Kb PICO-RAM macro occupies 0.074 mm$^2$, where the memory array, vertical/horizontal drivers, and ADC take 70.9\%, 14.7\%, and 4.6\% of the total area, respectively. The DAC area is negligible as it is embedded into the array. In all experiments, the test chip is interfaced with a host PC through an FPGA.

\subsection{Linearity and Computing Accuracy}
All analog components in the computing path, including DAC, analog MAC, analog shift-and-add, and ADC, contribute to the nonidealities of the system. We thoroughly evaluate all components to prove their superb linearity and accuracy.

   \begin{figure}[t]
      \centering
      \includegraphics[scale=1.05]{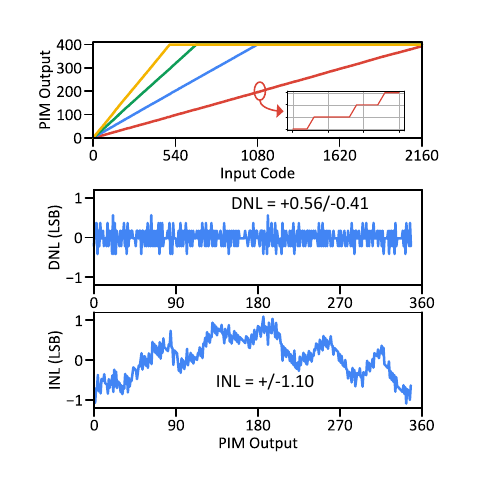}
      \caption{Measured end-to-end transfer curves at different gains and DNL/INL performance with gain = 1.}
      \label{linearity_input}
    \end{figure}

\textbf{Linearity with input sweeping.} In this linearity measurement, we store all `1's in the SRAM and sweep the inputs from 0 to maximum, i.e. $15\times144=2160$. Therefore, nonlinearities from DAC, analog MAC, and ADC are included. For a typical 8.5-bit CIM MVM group without any calibration, DNL and INL are bounded between +0.56/-0.41 and +/-1.10 LSB, respectively, as shown in Fig.~\ref{linearity_input}. The major error comes from the TD-ADC due to the restricted area for layout matching. By tuning the reference current in the VTC, the analog computing voltage can be amplified with a gain of up to 4 while maintaining satisfactory linearity. Due to the sparsity of DNN models and the effect of the Central Limit Theorem,  activations typically stay within a portion of the full dynamic range. Therefore, providing this gain effectively reduces quantization error.
\begin{figure}
\includegraphics[scale=1.1]{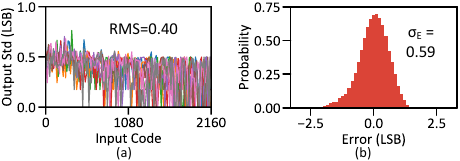}
    \centering
      \caption{Measured (a) standard deviation of output codes under thermal noise across all input codes for 8 CIM MVM groups; (b) computing error distribution including both nonlinearity and random noise.}
      \label{noise}
      \end{figure} 
      
   \begin{figure}[t]
       \centering
      \includegraphics[scale=0.35]{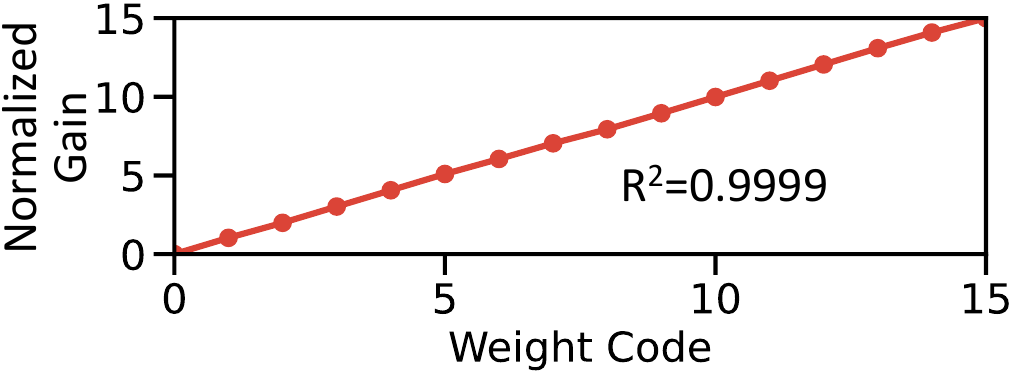}
      \caption{Measured gain of transfer curves with different weight configurations.}
      \vspace{-2ex}
      \label{linearity_weight}
   \end{figure}

\textbf{Random errors and error distribution.} With input sweeping and each code repeating 50 times, we further characterize the influence of thermal noise. Fig.~\ref{noise}(a) shows the root-mean-square (RMS) standard deviation of CIM outputs across all input codes, with an average of 0.4 LSB across 8 CIM MVM groups in a macro. This noise level is sufficient for systems targeting low power and small areas, yet can be further improved with a larger capacitor value, a less noisy RO, and a lower-noise zero detector. Considering both random errors and nonlinearity, the computation error distribution in Fig.~\ref{noise}(b) shows a standard deviation ($\sigma_E$) of 0.59 LSB.

\textbf{Linearity with weight sweeping.} Here, we evaluate the linearity of analog shift-and-add circuits. All 4-bit weights in SRAM are programmed to the same value. For each possible weight value, we sweep the input to obtain a transfer curve and calculate its slope. Ideally, the slope of the curve should linearly increase with the weight value. Fig.~\ref{linearity_weight} plots the slopes (gain) of all 16 transfer curves, showing consistent steps between neighboring codes.
The largest error happens at code `1000', where three bits are flipped from the last code `0111'. Despite the almost perfect capacitor matching, the error still exists because of the parasitic capacitors from the additional separation switches, prechargers, and P-sum combiners connected to the MAC Line. 

\subsection{PVT Robustness}

   \begin{figure*}[t]
      \centering
      \includegraphics[scale=1.02]{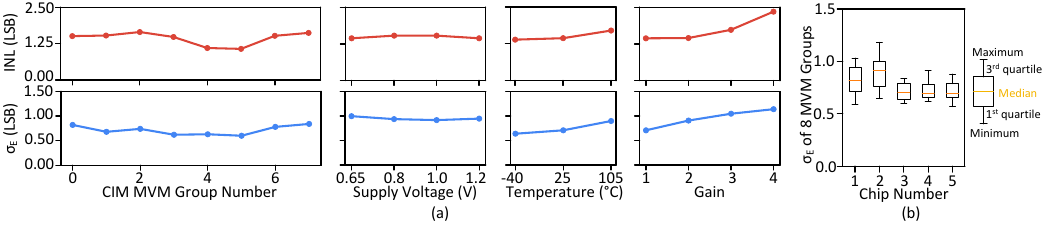}
      \vspace{-1ex}
      \caption{Measured end-to-end error distribution and INL (1) under PVT variations and different gains (2) across 5 chips.}
      \label{pvt}
   \end{figure*} 
   

   \begin{figure*}[t]
      \centering
      \includegraphics[scale=1.3]{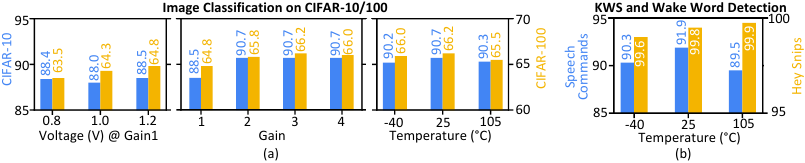}
       \vspace{-1ex}
      \caption{Inference accuracy on (a) CIFAR-10/100 across voltages, temperatures, and transfer curve gains; (b) Speech Commands/Hey Snips across temperatures.}
      \label{inference}
   \end{figure*} 

Based solely on passive components, the proposed fully capacitive CIM operation achieves superior tolerance of PVT variations. The highly digital TD-ADC also has great scalability to voltage. 
As shown in Fig.~\ref{pvt}(a), we examine the process variation by measuring $\sigma_E$ and INL of 8 CIM MVM groups in a single macro, where the difference between the best and worse ones is only 0.24 and 0.58 LSB, respectively. Furthermore, Fig.~\ref{pvt}(b) evaluates $\sigma_E$ across 5 chips, showing the similar distribution of $\sigma_E$ across 8 MVM groups in each chip. Similarly, we evaluate $\sigma_E$ and INL across 0.65 to 1.2 V and -40 to 105 \textdegree C in Fig.~\ref{pvt}(a), proving the robustness over voltage and temperature variations. This is so far the widest operation range of voltages and temperatures reported among CIM designs. Note that at 0.65 V, the resolution of CIM degrades to 8 bits due to the reduction of ADC input range, yet it still maintains satisfactory computing accuracy.

In addition to PVT variations, we also examine the computing accuracy under different gains when tuning the reference current. Theoretically, a smaller reference current results in a greater gain and a smaller quantization error, but also incurs more noise in the current source. As shown in Fig.~\ref{pvt}(a), the $\sigma_E$ and INL scale much slower than the gain, which proves the benefits of reduction in quantization errors outweigh the incurred nonidealities.

\subsection{Deep Learning Inference}

\textbf{Image classification on CIFAR-10/100.} A 4-bit quantized ResNet-20 is deployed for the inference on CIFAR-10 and CIFAR-100 datasets. To map the model into the marco, the 3$\times$3 3-D filters are unrolled into 1-D vectors and stored in one row of the macro with up to 16 input channels. As the model does not fit into a single macro, reloading the memory is necessary to complete the inference. We conduct a CIM-aware model training~\cite{jin2022pim}, where the additional ADC quantization step is considered during the forward propagation. Thanks to the bit-parallel computation, the training speed approaches that of the standard training method, whereas modeling bit-serial CIM in our training framework takes up to $3.7\times$ more time on an NVIDIA A-10 GPU because of the extra tensor dimensions. Meanwhile, bit-serial CIM frequently causes convergence failures due to the approximated bit-wise gradient. At the gain of 3, the system achieves 90.7\% and 66.2\% inference accuracy on CIFAR-10 and CIFAR-100 (see Fig.~\ref{inference}(a)), respectively, which is 0.3\% and 0.1\% less than the software emulation without analog nonidealities. It is worth mentioning that the custom training flow does not include modeling the nonidealities of individual chips and therefore is scalable to real-world industry production.

We evaluate the inference accuracy across different voltages and temperatures, as shown in Fig.~\ref{inference}(a). It drops 0.5\% on CIFAR-10 when the supply is 1.0 V and 1.3\% on CIFAR-100 when the supply is 0.8 V. For temperature variation, the accuracy reduces by 1.0\% on CIFAR-10 at -40 \textdegree C and 0.7\% on CIFAR-100 at 105 \textdegree C. Due to additional nonidealities, slight accuracy degradation is expected with variations, yet the system generally maintains reliable computing accuracy across a wide range of operating conditions.

    \begin{figure}[t]
      \centering
      \includegraphics[scale=1.1]{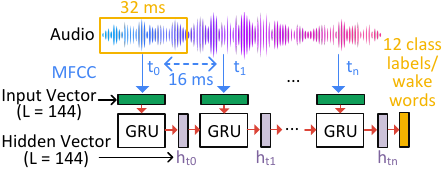}
       \vspace{-1ex}
      \caption{Custom RNN architecture for KWS and wake word detection.}
      \label{model}
      \vspace{3ex}
      \includegraphics[scale=1.1]{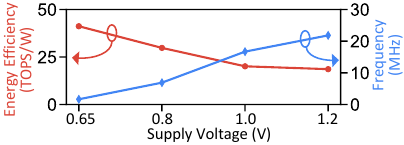}
       \vspace{-1ex}
      \caption{Measured energy efficiency and clock frequency from 0.65 to 1.2 V.}
      \vspace{-1ex}
      \label{voltage}
   \end{figure} 

\textbf{Keyword spotting (KWS) and wake word detection.} Although prior studies primarily focus on CNN acceleration, CIM SRAM is exceptionally conducive to recurrent neural networks (RNNs) that typically have greater matrix dimensions. According to~\cite{vermaInMemoryComputingAdvances2019}, with a large dimension of MVM, more SRAM cells will participate in the analog computation to amortize the energy from bitline/wordline and data converters. More importantly, the data-intensive RNN applications require larger on-chip memory, particularly suitable for our clustering structure with enhanced storage density. We customize a 0.16-M parameter 4-bit quantized Gated Recurrent Unit (GRU) architecture where both input and hidden vectors have a dimension of 144 to perfectly fit into the SRAM, as illustrated by Fig.~\ref{model}. Each 32-ms segment from the streaming audio will first be encoded by mel-frequency cepstral coefficients (MFCC) to identify the human speech spectrum and then sent to the GRU for CIM-based MVM. We simultaneously conduct KWS and wake word detection on a single RNN, achieving 91.9\% and 99.9\% inference accuracy on Speech Commands and Hey Snips datasets~\cite{coucke2019efficient} (see Fig.~\ref{inference}(b)), respectively, which shows 1.1\% and $\sim 0$\% degradation than software ideal emulation. Fig.~\ref{inference}(b) also consolidates the robustness over temperature variations for the audio tasks.

\newcommand{\tabincell}[2]{\begin{tabular}{@{}#1@{}}#2\end{tabular}}
\begin{table*}[!t]
\caption{\textbf{Summary of This Work and Comparison with State-of-the-Art Analog CIM SRAMs for Multi-Bit MVM}}
\label{table}
\centering
\setlength{\tabcolsep}{3.6pt}
\renewcommand{\arraystretch}{1.5}
\begin{tabular}{|p{70pt}|p{40pt}|p{40pt}|p{40pt}|p{40pt}|p{40pt}|p{40pt}|p{40pt}|p{40pt}|}
\hline
& 
\multicolumn{1}{c|}{\textbf{\centering\arraybackslash This Work}}&
\centering\arraybackslash \textbf{ISSCC$'$23\cite{hsieh2023a70}}&
\centering\arraybackslash \textbf{JSSC$'$23\cite{wangChargeDomainSRAM2023}}& 
\centering\arraybackslash \textbf{ISSCC$'$20\cite{dong15351TOPS3722020}}& 
\centering\arraybackslash \textbf{JSSC$'$21\cite{jiaScalableProgrammableNeural2021}}&
\centering\arraybackslash \textbf{JSSC$'$21\cite{capram}}&
\centering\arraybackslash \textbf{JSSC$'$22\cite{9896828}}&
\centering\arraybackslash \textbf{ISSCC$'$23\cite{chen2023a22nm}}
\\\hline

\multicolumn{1}{|c|}{\textbf{Technology (nm)}} 
&\multicolumn{1}{c|}{\textbf{65}}
&\multicolumn{1}{c|}{12} 
&\multicolumn{1}{c|}{22}
& \multicolumn{1}{c|}{7} 
&\multicolumn{1}{c|}{16}
&\multicolumn{1}{c|}{65}
&\multicolumn{1}{c|}{28}
&\multicolumn{1}{c|}{22}
\\\hline

\multicolumn{1}{|c|}{\textbf{Memory Capacity}} & 
\multicolumn{1}{c|}{\textbf{40.5 Kb}}& 
\multicolumn{1}{c|}{128 Kb}&
\multicolumn{1}{c|}{64 Kb}&
\multicolumn{1}{c|}{4 Kb}& 
\multicolumn{1}{c|}{281 Kb$\times$16}&
\multicolumn{1}{c|}{64 Kb}&
\multicolumn{1}{c|}{96 Kb$\times$4}&
\multicolumn{1}{c|}{64 Kb$\times$2}
\\\hline

\multicolumn{1}{|c|}{\textbf{Analog MAC Precision}} 
& \multicolumn{1}{c|}{\textbf{4b in$\times$4b w}}
& \multicolumn{1}{c|}{8b in$\times$8b w} 
& \multicolumn{1}{c|}{8b in$\times$8b w}
& \multicolumn{1}{c|}{4b in$\times$4b w} 
& \multicolumn{1}{c|}{1b in$\times$1b w}
& \multicolumn{1}{c|}{4b in$\times$1b w}
& \multicolumn{1}{c|}{2b in$\times$1b w}
& \multicolumn{1}{c|}{1b in$\times$1b w}

\\\hline

\multicolumn{1}{|c|}{\textbf{{Multi-Bit  Scheme}}}
& \multicolumn{1}{c|}{\textbf{{BP}}}
& \multicolumn{1}{c|}{{BP}} 
& \multicolumn{1}{c|}{{BP}}
& \multicolumn{1}{c|}{{BP}} 
& \multicolumn{1}{c|}{{BS}}
& \multicolumn{1}{c|}{{WBS}}
& \multicolumn{1}{c|}{{WBS}}
& \multicolumn{1}{c|}{{BS}}

\\\hline

\multicolumn{1}{|c|}{\textbf{Memory Density (Kb/mm$^2$)}}  
& \multicolumn{1}{c|}{\textbf{559 }}
& \multicolumn{1}{c|}{N/A}
& \multicolumn{1}{c|}{512 }
& \multicolumn{1}{c|}{1250 }
& \multicolumn{1}{c|}{465}
& \multicolumn{1}{c|}{366}
& \multicolumn{1}{c|}{840 }
& \multicolumn{1}{c|}{312}
\\\hline

\multicolumn{1}{|c|}{\begin{tabular}{@{}c@{}}\textbf{Memory Density}\\\textbf{(Normalized to 65 nm)$^\mathrm{a}$}\end{tabular} }
& \multicolumn{1}{c|}{\textbf{559 }}

& \multicolumn{1}{c|}{N/A}
& \multicolumn{1}{c|}{59 }
& \multicolumn{1}{c|}{14 }
& \multicolumn{1}{c|}{28 }
& \multicolumn{1}{c|}{366}
& \multicolumn{1}{c|}{156}
& \multicolumn{1}{c|}{36}

\\\hline

\multicolumn{1}{|c|}{\textbf{ADC Resolution}} 
& \multicolumn{1}{c|}{\textbf{8.5}} 
& \multicolumn{1}{c|}{8}
& \multicolumn{1}{c|}{8}
& \multicolumn{1}{c|}{4}
& \multicolumn{1}{c|}{8}
& \multicolumn{1}{c|}{6} 
& \multicolumn{1}{c|}{5} 
& \multicolumn{1}{c|}{7} 
 \\\hline

 \multicolumn{1}{|c|}{{\textbf{Error Distribution $\sigma$ (LSB)}}} 
& \multicolumn{1}{c|}{\textbf{{0.59}}} 
& \multicolumn{1}{c|}{{N/A}}
& \multicolumn{1}{c|}{{2.27}}
& \multicolumn{1}{c|}{{N/A}}
& \multicolumn{1}{c|}{{N/A}}
& \multicolumn{1}{c|}{{1.34}} 
& \multicolumn{1}{c|}{{N/A}} 
& \multicolumn{1}{c|}{{N/A}} 
 \\\hline

\multicolumn{1}{|c|}{\textbf{Supply Voltage (V)}}  
& \multicolumn{1}{c|}{\textbf{0.65$\sim$1.2}} 
& \multicolumn{1}{c|}{0.5$\sim$0.85}
& \multicolumn{1}{c|}{0.7$\sim$1.1}
& \multicolumn{1}{c|}{0.65$\sim$1}
& \multicolumn{1}{c|}{0.8}
& \multicolumn{1}{c|}{1.2}
& \multicolumn{1}{c|}{0.7$\sim$0.9}
& \multicolumn{1}{c|}{0.72$\sim$0.82}
\\\hline

\multicolumn{1}{|c|}{\textbf{Temperature (°C)}}  
& \multicolumn{1}{c|}{\textbf{-40$\sim$105}} 
& \multicolumn{1}{c|}{25$\sim$60}
& \multicolumn{1}{c|}{-20$\sim$70}
& \multicolumn{1}{c|}{N/A}
& \multicolumn{1}{c|}{N/A}
& \multicolumn{1}{c|}{N/A}
& \multicolumn{1}{c|}{N/A}
& \multicolumn{1}{c|}{N/A}
\\\hline

\multicolumn{1}{|c|}{\textbf{Computing Parallelism$^\mathrm{b}$}}
& \multicolumn{1}{c|}{\textbf{144}}
& \multicolumn{1}{c|}{64} 
& \multicolumn{1}{c|}{64}
& \multicolumn{1}{c|}{64}
& \multicolumn{1}{c|}{1152}
& \multicolumn{1}{c|}{128}
& \multicolumn{1}{c|}{16}
& \multicolumn{1}{c|}{64}
\\\hline

\multicolumn{1}{|c|}{\textbf{\begin{tabular}{@{}c@{}}\textbf{Throughput}\\\textbf{(GOPS)}\end{tabular}}}
& \multicolumn{1}{c|}{\begin{tabular}{@{}c@{}}\textbf{3.8@0.65V}\\ \textbf{50.3@1.2V}\end{tabular}} 
& \multicolumn{1}{c|}{1024}
& \multicolumn{1}{c|}{\begin{tabular}{@{}c@{}}600@0.7V\\1000@1.1V\end{tabular}}
& \multicolumn{1}{c|}{\begin{tabular}{@{}c@{}}372.4@0.8V\\455.1@1.0V\end{tabular}}
& \multicolumn{1}{c|}{3000}
& \multicolumn{1}{c|}{573}
& \multicolumn{1}{c|}{7675}
& \multicolumn{1}{c|}{490$\sim$600}
\\\hline

\multicolumn{1}{|c|}{\textbf{\begin{tabular}{@{}c@{}}\textbf{Bitwise}\\\textbf{Throughput$^\mathrm{c}$}\end{tabular}}}
& \multicolumn{1}{c|}{\begin{tabular}{@{}c@{}}\textbf{60.8@0.65V}\\ \textbf{804.8@1.2V}\end{tabular}} 
& \multicolumn{1}{c|}{65536}
& \multicolumn{1}{c|}{\begin{tabular}{@{}c@{}}38400@0.7V\\64000@1.1V\end{tabular}}
& \multicolumn{1}{c|}{\begin{tabular}{@{}c@{}}5958.4@0.8V\\7281.1@1.0V\end{tabular}}
& \multicolumn{1}{c|}{48000}
& \multicolumn{1}{c|}{2294}
& \multicolumn{1}{c|}{122803}
& \multicolumn{1}{c|}{\begin{tabular}{@{}c@{}}31360$\sim$\\38400\end{tabular}}
\\\hline

\multicolumn{1}{|c|}{\begin{tabular}{@{}c@{}}\textbf{Compute Density}\\\textbf{(TOPS/$\bm{m m^{2}}$)}\end{tabular}}
& \multicolumn{1}{c|}{\begin{tabular}{@{}c@{}}\textbf{0.05@0.65V}\\ \textbf{0.68@1.2V}\end{tabular}} 
& \multicolumn{1}{c|}{N/A}
& \multicolumn{1}{c|}{\begin{tabular}{@{}c@{}}2.4@0.7V\\4.0@1.1V\end{tabular}}
& \multicolumn{1}{c|}{\begin{tabular}{@{}c@{}}116.4@0.8V\\142. @1.0V\end{tabular}}
& \multicolumn{1}{c|}{42.72}
& \multicolumn{1}{c|}{3.4}
& \multicolumn{1}{c|}{1.64}
& \multicolumn{1}{c|}{1.19$\sim$1.44}
\\\hline

\multicolumn{1}{|c|}{\begin{tabular}{@{}c@{}}\textbf{Bitwise Compute Density$^\mathrm{c}$}\\\textbf{(Normalized to 65 nm)$^\mathrm{b}$}\end{tabular}}
& \multicolumn{1}{c|}{\begin{tabular}{@{}c@{}}\textbf{0.8@0.65V}\\ \textbf{10.88@1.2V}\end{tabular}} 
& \multicolumn{1}{c|}{N/A}
& \multicolumn{1}{c|}{\begin{tabular}{@{}c@{}}17.5@0.7V\\29.3@1.1V\end{tabular}}
& \multicolumn{1}{c|}{\begin{tabular}{@{}c@{}}21.5@0.8V\\26.3@1.0V\end{tabular}}
& \multicolumn{1}{c|}{2.58}
& \multicolumn{1}{c|}{13.6}
& \multicolumn{1}{c|}{4.8}
& \multicolumn{1}{c|}{8.72$\sim$10.55}
\\\hline

\multicolumn{1}{|c|}{\begin{tabular}{@{}c@{}}\textbf{Energy Efficiency}\\\textbf{(TOPS/W)}\end{tabular}}
& \multicolumn{1}{c|}{\begin{tabular}{@{}c@{}}\textbf{40.2@0.65V}\\ \textbf{18.6@1.2V}\end{tabular}} 
& \multicolumn{1}{c|}{70.85}
& \multicolumn{1}{c|}{\begin{tabular}{@{}c@{}}32.2@0.7V\\15.5@1.1V\end{tabular}}
& \multicolumn{1}{c|}{\begin{tabular}{@{}c@{}}351@0.8V\\321@1.0V\end{tabular}}
& \multicolumn{1}{c|}{1936}
& \multicolumn{1}{c|}{49.4}
& \multicolumn{1}{c|}{60.28$\sim$94.31}
& \multicolumn{1}{c|}{16.02$\sim$21.38}
\\\hline

\multicolumn{1}{|c|}{\begin{tabular}{@{}c@{}}\textbf{Bitwise Energy Efficiency$^\mathrm{c}$}\\\textbf{(Normalized to 65 nm)$^\mathrm{d}$}\end{tabular}}
& \multicolumn{1}{c|}{\begin{tabular}{@{}c@{}}\textbf{643.2@0.65V}\\ \textbf{297.6@1.2V}\end{tabular}} 

& \multicolumn{1}{c|}{154.5}
& \multicolumn{1}{c|}{\begin{tabular}{@{}c@{}}236.1@0.7V\\113.6@1.1V\end{tabular}}
& \multicolumn{1}{c|}{\begin{tabular}{@{}c@{}}65.1@0.8V\\59.6@1.0V\end{tabular}}
& \multicolumn{1}{c|}{117.3}
& \multicolumn{1}{c|}{197.6}
& \multicolumn{1}{c|}{179.0$\sim$280.0}
& \multicolumn{1}{c|}{117.5$\sim$156.74}
\\\hline

\multicolumn{1}{|c|}{\textbf{Tested Network Models}}
& \multicolumn{1}{c|}{\begin{tabular}{@{}c@{}}\textbf{ResNet-20}\\ \textbf{RNN}\\ \textbf{{(4 bit)}}\end{tabular}} 
& \multicolumn{1}{c|}{\begin{tabular}{@{}c@{}}ResNet-20\\MobileNet\\{(8 bit)}\end{tabular}}
& \multicolumn{1}{c|}{\begin{tabular}{@{}c@{}}MLP\\LeNet-5\\{(8 bit)}\end{tabular}}
& \multicolumn{1}{c|}{N/A}
& \multicolumn{1}{c|}{\begin{tabular}{@{}c@{}}VGG\\{(4 bit)}\end{tabular}}
& \multicolumn{1}{c|}{\begin{tabular}{@{}c@{}}LeNet-5\\ResNet-20\\{(4 bit)}\end{tabular}}
& \multicolumn{1}{c|}{\begin{tabular}{@{}c@{}}ResNet-20\\{(4 bit)}\end{tabular}}
& \multicolumn{1}{c|}{\begin{tabular}{@{}c@{}}ResNet-20\\{(8 bit)}\end{tabular}}
\\\hline

\multicolumn{1}{|c|}{\textbf{Datasets}}
& \multicolumn{1}{c|}{\begin{tabular}{@{}c@{}c@{}c@{}}\textbf{CIFAR-10}\\\textbf{CIFAR-100}\\\textbf{Speech Com.} \\\textbf{Hey Snips}\end{tabular}} 
& \multicolumn{1}{c|}{\begin{tabular}{@{}c@{}}CIFAR-100\\VWW\end{tabular}}
& \multicolumn{1}{c|}{MNIST}
& \multicolumn{1}{c|}{N/A}
& \multicolumn{1}{c|}{\begin{tabular}{@{}c@{}}CIFAR-10\\ImageNet\end{tabular}}
& \multicolumn{1}{c|}{\begin{tabular}{@{}c@{}}MNIST\\CIFAR-10\end{tabular}}
& \multicolumn{1}{c|}{\begin{tabular}{@{}c@{}}CIFAR-10\\CIFAR-100\end{tabular}}
& \multicolumn{1}{c|}{CIFAR-10}
\\\hline

\multicolumn{1}{|c|}{\textbf{Inference Accuracy}}
& \multicolumn{1}{c|}{\begin{tabular}{@{}c@{}c@{}c@{}}\textbf{90.7\%}\\\textbf{66.2\%}\\\textbf{91.9\%}\\\textbf{99.9\%}\end{tabular}} 
& \multicolumn{1}{c|}{\begin{tabular}{@{}c@{}}67.8\%\\80.0\%\end{tabular}}
& \multicolumn{1}{c|}{\begin{tabular}{@{}c@{}}98.14\%\\96.85\%\end{tabular}}
& \multicolumn{1}{c|}{N/A}
& \multicolumn{1}{c|}{\begin{tabular}{@{}c@{}}91.52\%\\73.33\%\end{tabular}}
& \multicolumn{1}{c|}{\begin{tabular}{@{}c@{}}98.8\%\\89.0\%\end{tabular}}
& \multicolumn{1}{c|}{\begin{tabular}{@{}c@{}}91.21/91.85\%\\67.26/67.97\%\end{tabular}}
& \multicolumn{1}{c|}{91.95\%}
\\\hline

\multicolumn{9}{p{500pt}}{
{$^\mathrm{a}$ Assume area $\propto$ (technology)$^2$~\cite{yan20221}.\:
$^\mathrm{b}$ Measured as the number of rows accessed concurrently per analog MVM.
}}

\\
\multicolumn{8}{p{420pt}}{$^\mathrm{c}$ Bitwise Metric = Metric$\times$CIM Input Bitwidth$\times$CIM Weight Bitwidth. \:
$^\mathrm{d}$ Assume energy $\propto$ (technology)$^2$~\cite{10299656, yueSTICKERIM65Nm2022, sehgal2023compute, 10121238, biswas2019conv}.}

\\

\end{tabular}
\label{tab2}
\vskip -2.5ex
\end{table*}

\subsection{Area, Energy and Throughput}
The PICO-RAM macro with a fully thin-cell layout achieves a memory density of 559 Kb/mm$^2$, which is the highest among all CIM designs (normalized to 65 nm). Including all the extra area for CIM, the memory density of our CIM macro is only 31\% lower than a logic-rule 6T SRAM, similar to that of an 8T SRAM. It achieves 3.6$\times$ memory density than \cite{9896828} even though \cite{9896828} utilizes push-rule 6T cells and has 32 cells in a cluster. In practice, this is especially beneficial to CIM systems targeting fully on-chip weight storage for medium-sized models in ultra-low-power edge devices.

Fig.~\ref{voltage} demonstrates the 0.65-1.2 V system operates from 2 MHz to 22 MHz, achieving 40.2 TOPS/W at 0.65 V and 0.68 TOPS/mm$^2$ at 1.2 V with 4-bit input and 4-bit weights. There exists a trade-off between efficiency and memory density: the more cells in a cluster, the longer the metal wires and thus the more energy consumption and less computing parallelism. Although the system is not optimized for maximum efficiency given the 9-bit cluster structure, it achieves comparable efficiency as state-of-the-arts, thanks to the voltage scalability and ADC energy reduction techniques. It is worth noting that the cluster size is scalable, which means one can always include only one SRAM cell in a cluster to achieve the best efficiency while maintaining the uniform thin-cell layout. The detailed performance comparison is summarized in Table~\ref{table}. Since the performance of CIM macros usually scales with bit precision, throughput and efficiency metrics are normalized to 1-bit operations for a fair comparison among different designs, similar to~\cite{si1528nm64Kb2020, capram, wangChargeDomainSRAM2023, jiaScalableProgrammableNeural2021, sehgal2023compute}.

\section{Conclusion}
\label{sec:conclusion}
In conclusion, this paper presents PICO-RAM, a PVT-resilient charge-domain CIM SRAM macro with a thin-cell-compatible layout and in-situ multi-bit Bit-Parallel MVM computation. The DAC reuses the local MOM capacitors as the reference generator and directly samples the voltage on them for subsequent capacitive MAC operations. Further, the novel in-situ shift-and-add circuits reuse the same set of MOM capacitors in the array for inter-row charge-sharing, leading to a remarkable area reduction and enhanced linearity. All analog components in the array adopt a uniform 6T-thin-cell transistor layout, achieving the densest storage density among CIM designs. The ultra-compact TD-ADC utilizes a dual-threshold voltage-to-time conversion, leading to 55.8\% energy reduction compared to conventional designs. A 65-nm prototype demonstrates excellent computing accuracy and robust operation across 0.65 V to 1.2 V and -40 to 105~\textdegree C. End-to-end machine learning inference tasks, including image classification and audio detection, are conducted with multiple supply voltage and temperature conditions. With 4-bit quantized ResNet-20 and a custom RNN, it achieves 90.7\%, 66.2\%, 91.9\%, and 99.9\% accuracy on CIFAR-10, CIFAR-100, Speech Commands, and Hey Snips datasets, respectively. The system achieves 40.9 TOPS/W energy efficiency and 0.65 TOPS/mm$^2$ with 4-bit activations and 4-bit weights.


%

\appendices




\ifCLASSOPTIONcaptionsoff
  \newpage
\fi



\bibliography{bibtex/bib/all.bib}
\bibliographystyle{ieeetr}
%



%

\begin{IEEEbiography}[{\includegraphics[width=1in,height=1.25in,clip,keepaspectratio]{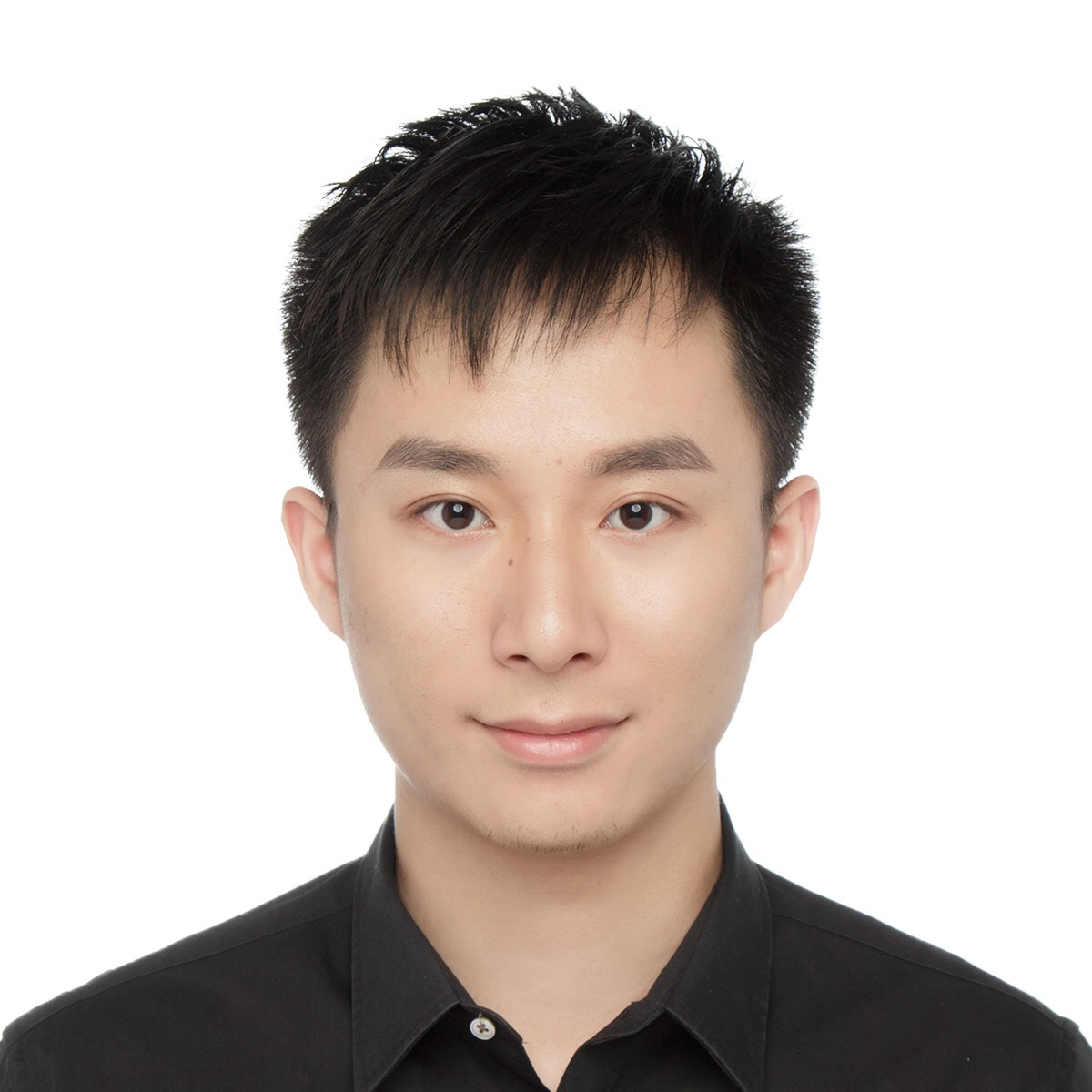}}]{Zhiyu Chen}  (Member, IEEE) received his B.E. degree in Electrical Engineering from Nanjing University, Nanjing, China, in 2018, and his Ph.D. degree in Electrical and Computer Engineering from Rice University, Houston, TX, in 2024. He is currently an SRAM design engineer at Apple Inc., Cupertino, CA.

His research focuses on mixed-signal non-Von Neumann accelerators for machine learning.
\end{IEEEbiography}

\begin{IEEEbiography}
[{\includegraphics[width=1in,height=1.25in,clip,keepaspectratio]{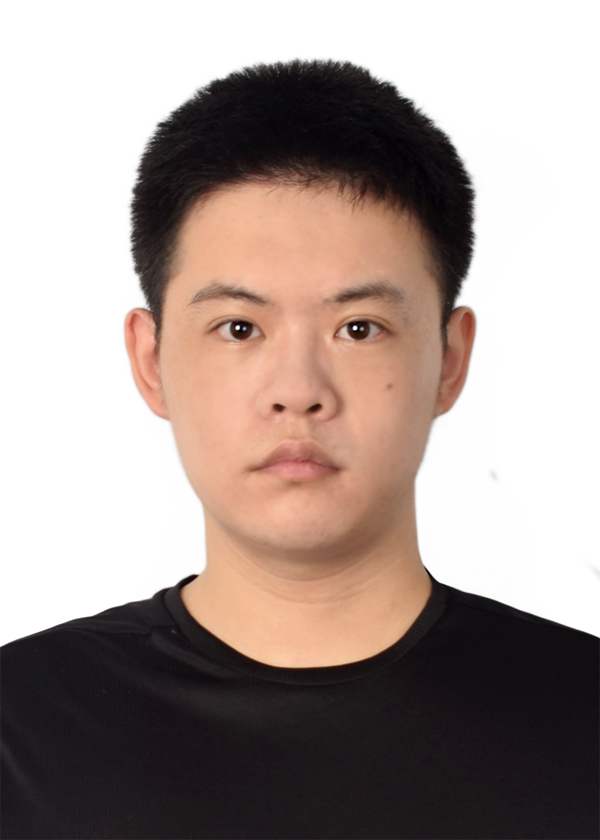}}]{Ziyuan Wen} (Graduate Student Member, IEEE) received his bachelor's degree in Optical and Electronic Information from
the Huazhong University of Science and Technology, Wuhan, China in 2022. He is currently
pursuing the Ph.D. degree in Electrical and Computer Engineering at Rice University, Houston, TX.

His research interests include low-power integrated circuits for streaming data
processors, biomedical applications, and in-memory
computing accelerators for deep learning.
\end{IEEEbiography}

\begin{IEEEbiography}
[{\includegraphics[width=1in,height=1.25in,clip,keepaspectratio]{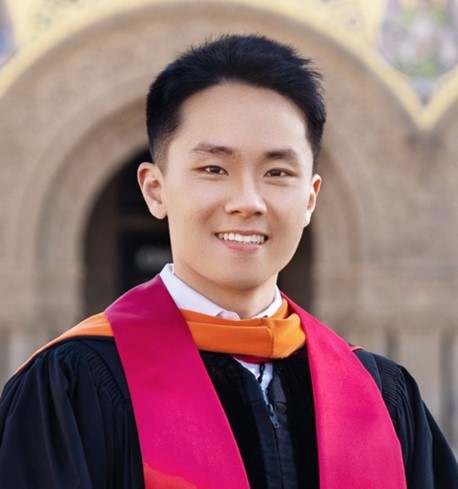}}]{Weier Wan} is heading the Enterprise Solutions group and is a founding member at Aizip, a Silicon Valley startup providing TinyML and edge AI software. He received his Ph.D. degree in Electrical Engineering from Stanford University in 2022, where he worked on designing compute-in-memory hardware to enable efficient edge intelligence. His research work has been published in top journals and conferences, including Nature, International Solid-State Circuits Conference (ISSCC), and Symposium on VLSI Technology and Circuits. Previously, he received his master’s degree in Electrical Engineering from Stanford University in 2018 and his bachelor’s degree in Physics, Electrical Engineering, and Computer Sciences from the University of California, Berkeley in 2015.
\end{IEEEbiography}

\begin{IEEEbiography}
[{\includegraphics[width=1in,height=1.25in,clip,keepaspectratio]{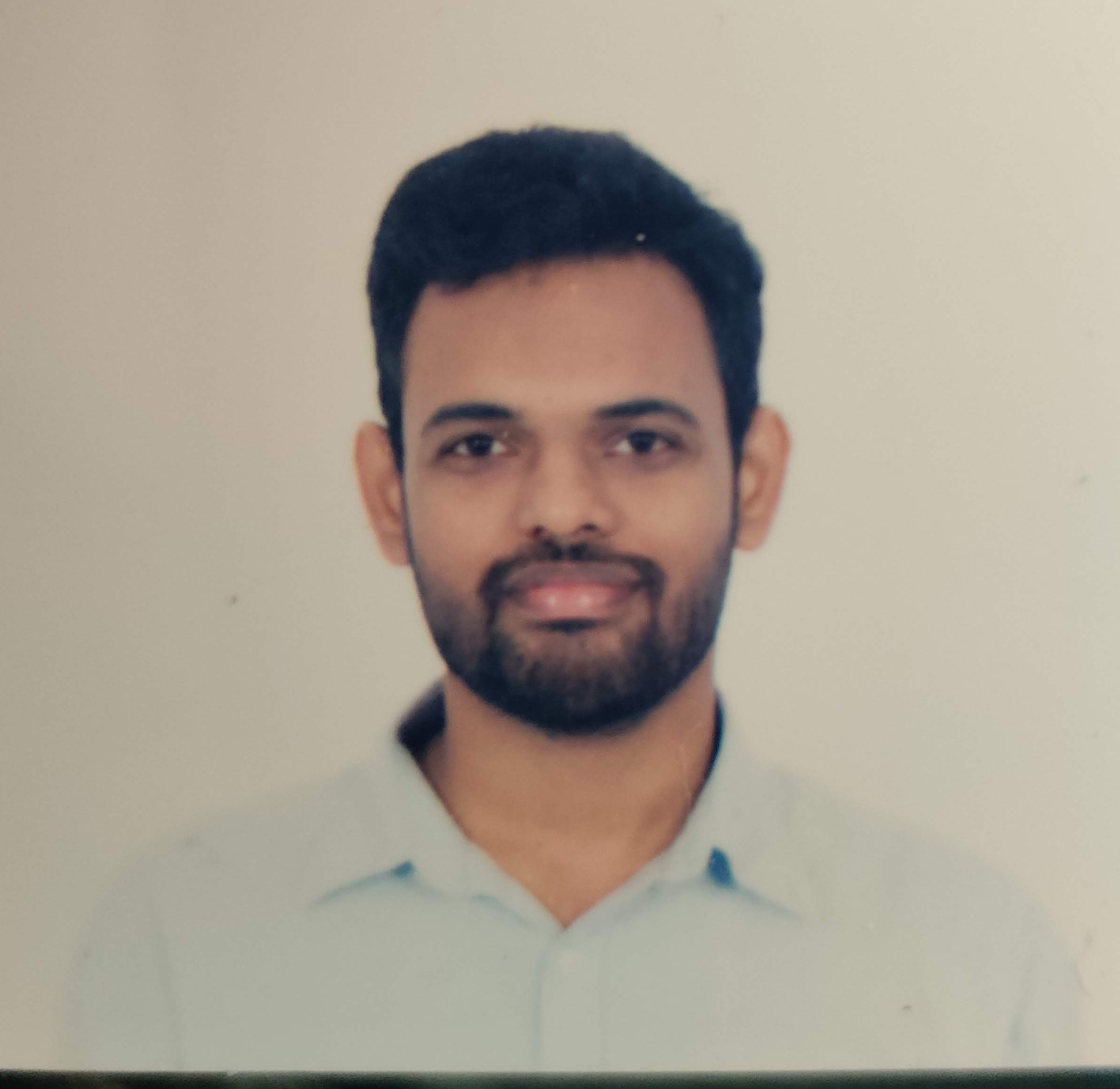}}] {Akhil Pakala} (Graduate Student Member, IEEE) received a dual degree (bachelor’s and master’s) from IIT Madras, Chennai, India, in 2019. He is currently working toward a master’s degree at Rice University, Houston, TX, USA. From 2019 to 2020, he worked with Samsung Semiconductor Research and Development, Bengaluru, India, on SerDes PHY IP. His current research interests include designing digital and mixed-signal circuits for security and machine-learning applications.
\end{IEEEbiography}

\begin{IEEEbiography}
[{\includegraphics[width=1in,height=1.25in,clip,keepaspectratio]{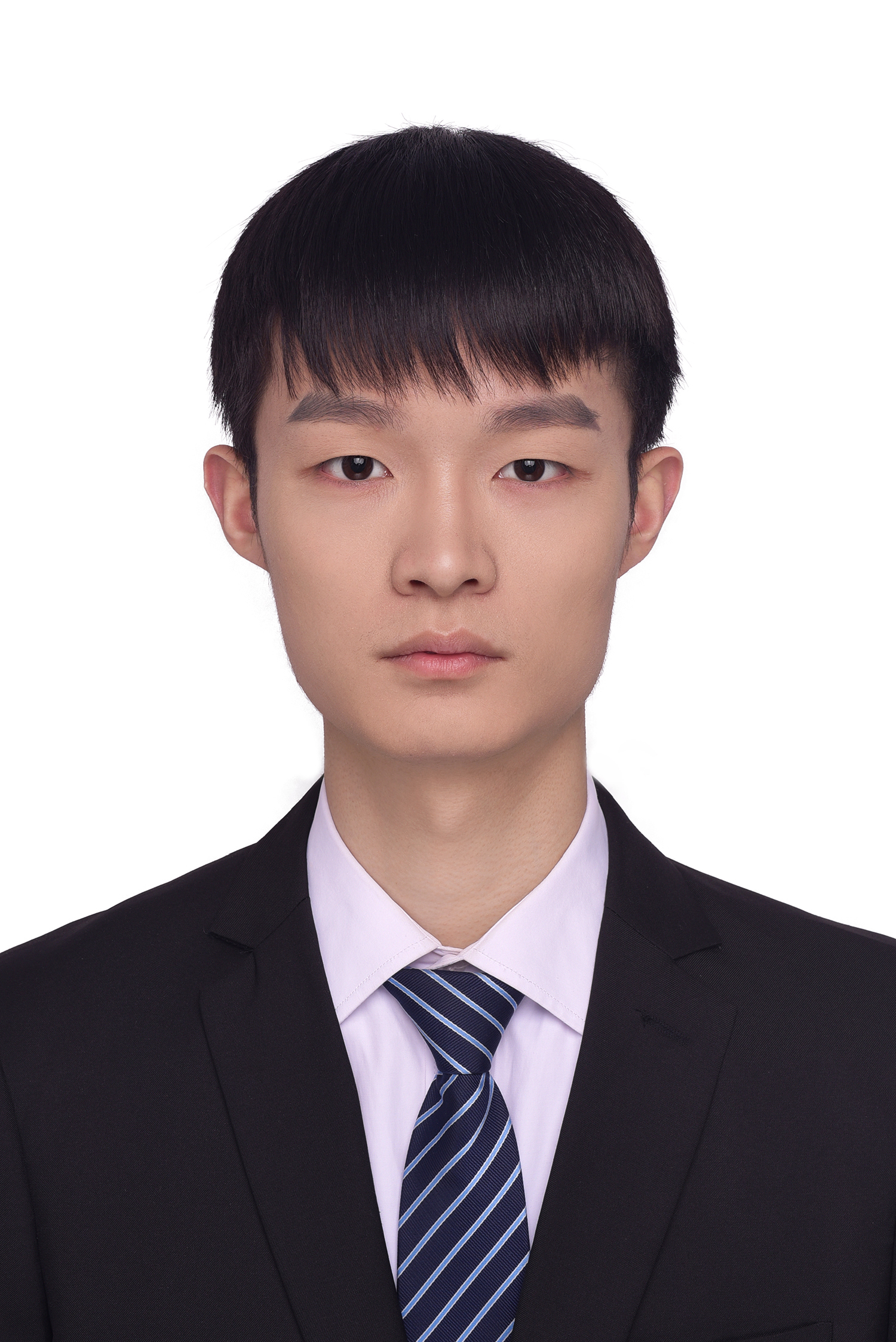}}] {Yiwei Zou} (Graduate Student Member, IEEE) received the B.E. degree in Integrated Circuits and Systems from Huazhong University of Science and Technology, Wuhan, China, in 2022. He is currently working toward his Ph.D. degree in Electrical and Computer Engineering at Rice University, Houston, TX. 

His research interests include analog and mixed-signal integrated circuits design for power management and bio-electronics.
\end{IEEEbiography}

\begin{IEEEbiography}
[{\includegraphics[width=1in,height=1.25in,clip,keepaspectratio]{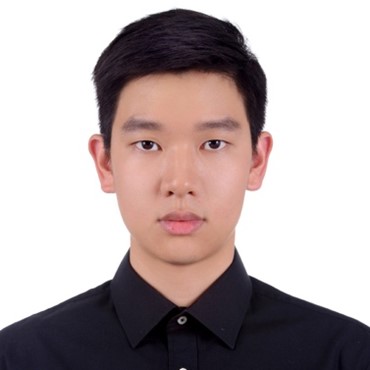}}] {Wei-Chen Wei} received the M.S. degree from the Institute of Electrical Engineering, National Tsing Hua University, Hsinchu, Taiwan, in 2019. He is currently pursuing a Ph.D. degree in Computer Engineering at Texas A\&M University, College Station, Texas. 

His current research interests include model compression algorithms for large language models and generative AI.

\end{IEEEbiography}

\begin{IEEEbiography}
[{\includegraphics[width=1in,height=1.25in,clip,keepaspectratio]{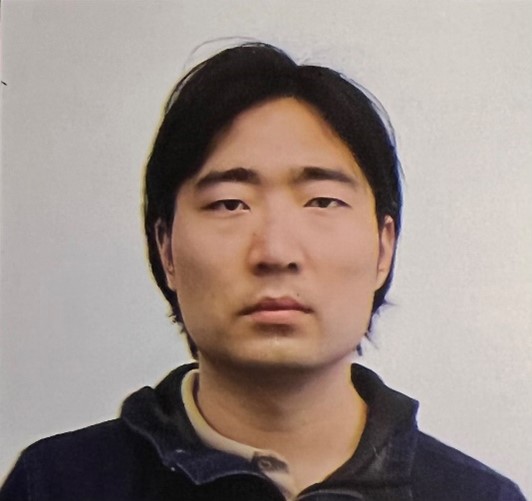}}] {Zengyi Li} received a bachelor's degree with a double major in Biophysics and Physiology \& Neuroscience from UC San Diego, California, USA, in 2017. He received his Ph.D. degree in Physics from UC Berkeley, California, USA, in 2022. Since then, he has been working as a research scientist in Aizip Inc., a US startup company providing tiny AI model solutions. His work focuses on developing audio AI models for various applications. 
\end{IEEEbiography}

\begin{IEEEbiography}
[{\includegraphics[width=1in,height=1.25in,clip,keepaspectratio]{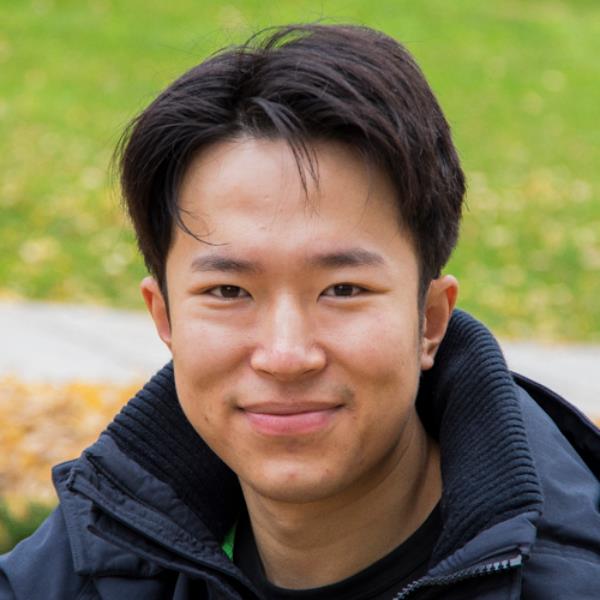}}] {Yubei Chen} is an Assistant Professor from the ECE department at UC Davis. He has worked with Professor Yann LeCun at Meta AI and NYU Center for Data Science as a postdoctoral researcher. Yubei received his MS/PhD in Electrical Engineering and Computer Sciences and MA in Mathematics at UC Berkeley under Professor Bruno Olshausen. His research interests span multiple aspects of representation learning. He explores the intersection of computational neuroscience and deep unsupervised learning, with the goal of improving our understanding of the computational principles governing unsupervised representation learning in both brains and machines, and reshaping our insights into natural signal statistics. He is a recipient of the NSF graduate fellowship, and ICLR Outstanding Paper Honorable Mention Award.
\end{IEEEbiography}

\begin{IEEEbiography}
[{\includegraphics[width=1in,height=1.25in,clip,keepaspectratio]{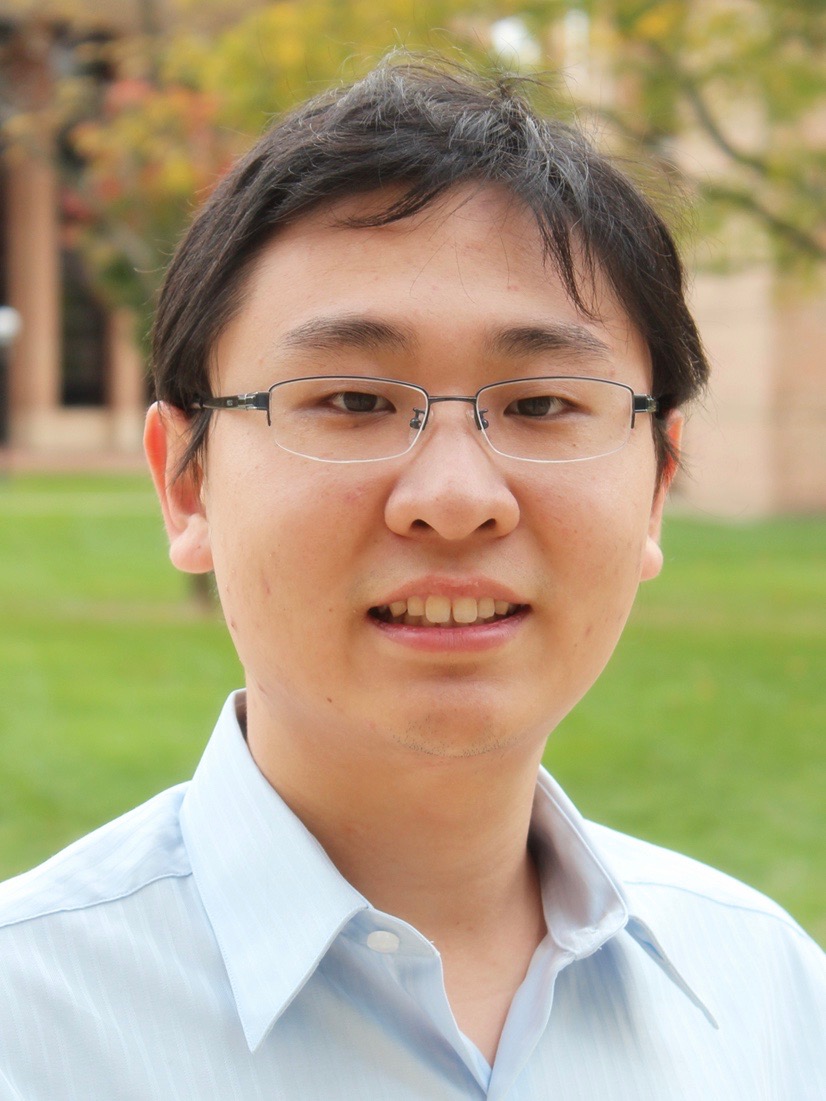}}] {Kaiyuan Yang} (Member, IEEE) is an Associate Professor of Electrical and Computer Engineering at Rice University, USA, where he leads the Secure and Intelligent Micro-Systems (SIMS) lab. He received his B.S. in Electronic Engineering from Tsinghua University, Beijing, China, in 2012, and his Ph.D. degree in Electrical Engineering from the University of Michigan, Ann Arbor, MI, in 2017. His research interests include low-power integrated circuits and system design for secure and intelligent microsystems, bioelectronics, hardware security, and mixed-signal computing. 

Dr. Yang is a recipient of 2022 National Science Foundation CAREER Award, 2016 IEEE Solid-State Circuits Society (SSCS) Predoctoral Achievement Award, and best paper awards from premier conferences across multiple fields, including 2022 ACM Annual International Conference on Mobile Computing and Networking (MobiCom), 2021 IEEE Custom Integrated Circuit Conference (CICC), 2016 IEEE International Symposium on Security and Privacy (Oakland), and 2015 IEEE International Symposium on Circuits and Systems (ISCAS), and several best paper award nominations. His research was also recognized as the research highlight at Communications of ACM (CACM) and ACM GetMobile magazines, the cover of Nature Biomedical Engineering, and IEEE Top Picks in Hardware and Embedded Security. He is currently serving as an associate editor of IEEE Transactions on VLSI Systems (TVLSI) and a technical program committee member of ISSCC, CICC, and DAC. 

\end{IEEEbiography}





\vfill


\end{document}